\documentclass[a4paper,11pt]{article}

\usepackage{amssymb,amsmath,amsfonts}
\usepackage{graphicx}
\usepackage{caption}
\usepackage{subcaption}
\usepackage{color}
\usepackage{hyperref}
\usepackage{cleveref}
\usepackage{bm}
\usepackage{listings}
\usepackage{multirow}
\usepackage[table]{xcolor}

\newcommand{\x}{\mathbf{x}}
\newcommand{\y}{\mathbf{y}}
\newcommand{\q}{\boldsymbol{\theta}}
\newcommand{\io}{\int_\Omega}
\newcommand{\ve}{\varepsilon}
\newcommand{\pa}{\partial}

\newcommand{\bge}[1]{{\color{black}{#1}}}
\newcommand{\ymp}[1]{{\color{black}{#1}}}

\newcommand{\ypre}[1]{{\color{black}{#1}}}
\newcommand{\ympre}[1]{{\color{black}{#1}}}
\newcommand{\ympree}[1]{{\color{black}{#1}}}
\newcommand{\be}[1]{{\color{black}{#1}}}

\begin{document}
\title{Scalar Reduction of a Neural Field Model with Spike Frequency Adaptation}
\author{Youngmin Park\footnote{Corresponding author. Email yop6@pitt.edu}\, \& Bard Ermentrout \\ Department of Mathematics\\University of Pittsburgh\\ Pittsburgh PA 15260}

\maketitle

\begin{abstract}
\ympre{We study a deterministic version of a \ympree{\be{one- and two-dimensional }attractor neural network model of hippocampal activity} first studied by Itskov et al 2011. We analyze the dynamics of the system on the ring and torus domain with an even periodized weight matrix, assuming weak and slow spike frequency adaptation and a weak stationary input current. On these domains, we find transitions from  \be{spatially localized stationary solutions (``bumps'')}  to \be{(periodically modulated)} solutions \be{(``sloshers'')}, as well as constant and non-constant velocity traveling \be{bumps} depending on the relative strength of \be{external} input current and adaptation. The weak and slow adaptation allows \be{for} a reduction of the system from a distributed partial integro-differential equation to a system of scalar \be{Volterra} integro-differential equations describing the movement of the centroid of the bump solution. Using this reduction, we show that on both domains, sloshing solutions arise through an Andronov-Hopf bifurcation and derive a normal form for the Hopf bifurcation on the ring. We also show existence and stability of constant velocity solutions on both domains using Evans functions. In contrast to existing studies, we assume a general weight matrix of Mexican-hat type in addition to a smooth firing rate function.}
\end{abstract}



\section{Introduction}
Spatially coherent activity states exist during normal brain function including mammalian path integration, head direction tracking, visual hallucination, working memory, spatial object location, and object orientation \cite{coombes_2005_biolCyb,coombes_bumps_2005,folias_bressloff_2005_prl}. Neural field models (also called continuous attractor neural networks) are one way to understand the mechanism underlying such spatially coherent phenomena \cite{fung_dynamical_2012,fung_spontaneous_2015,fung_fluctuation-response_2015}. In neural recordings and field models, these spatio-temporal dynamics manifest as traveling waves, spirals, or single or multiple localized ``bumps'' or ``pulses'' of activity \cite{coombes_2005_biolCyb}.

Extensive literature exists on the analysis of these behaviors. In particular, \cite{zhang_neural_2012} show the existence and stability of traveling bumps using multiple-layer neural fields. Several other studies use one of or a combination of short term depression and spike frequency adaptation. In \cite{kilpatrick_effects_2010}, the authors show that traveling pulses exist in a model with synaptic depression and adaptation when synaptic depression is sufficiently weak. For stronger synaptic depression, the traveling pulse ceases to exist via a saddle-node bifurcation. In \cite{fung_spontaneous_2015} the authors show that spontaneous motion of a bump solution exists for a neural field with only spike frequency adaptation, and in a similar neural field model with only short term synaptic depression. The authors in  \cite{pinto_ermentrout_2001_siam} show the existence of a traveling pulse solution in a neural field model with spike frequency adaptation. The previous two studies also show the existence of traveling wavefronts in their respective neural field models.

In addition to the analysis of traveling bumps or wavefronts, rich oscillatory solutions of neural fields are also possible. For example, with spatially localized input current and spike frequency adaptation, a bump solution may oscillate in diameter (breathers)\cite{bressloff_etal_2003_prl,folias_bressloff_2004_siam,folias_bressloff_2005_prl,folias_bressloff_2005_siam}, which may play a role in generating epileptiform activity \cite{folias_bressloff_2005_siam} and the processing of sensory stimuli \cite{folias_bressloff_2005_prl}. \ympre{There also exist studies of a combination of traveling and breathing pulses in an inhibitory-excitatory neural field \cite{folias2017traveling}}. In addition to breathers, there exist pulse-emitting neural fields \cite{kilpatrick_effects_2010,kilpatrick_spatially_2009}, oscillatory wavefronts \cite{bressloff_folias_2004_siam,bressloff_etal_2003_prl}, and spiral waves \cite{kilpatrick_spatially_2009}. 

Despite this large body of literature, the analyses \be{often} require particular assumptions. For example, \ympree{the existence of ``sloshing'' solutions -- bump solutions that oscillate periodically in the centroid -- that arise through a Hopf bifurcation is known under certain assumptions. \ymp{In early work, sloshers are shown to exist numerically using a rate model with a threshold nonlinearity \cite{hansel199813}}. In recent work, the authors of \cite{ermentrout2014spatiotemporal} show the existence of a Hopf bifurcation} with a cosine kernel and a particular choice of smooth firing rate function. In \cite{folias2011nonlinear}, Folias computes a normal form for the Hopf bifurcation using a general kernel, but \be{for} a Heaviside firing rate \be{function}.

Proving existence of other phenomena also require special assumptions. In \cite{faye2013existence}, the authors consider a neural field model on the real line with synaptic depression and prove the existence of a traveling pulse without a Heaviside assumption, but use the particular choice of a normalized exponential kernel. \ymp{In \cite{kilpatrick_pulse_2014}, the authors use a center manifold reduction to analyze the existence of moving bump solutions. They allow the firing rate to be sigmoidal or a Heaviside, but require a cosine kernel. Similar assumptions are made in \cite{}, where they assume a hyperbolic tangent firing rate function and a cosine kernel.}

The most general of such studies, \cite{pinto_ermentrout_2001_siam}, considers a neural field model on the real line with spike frequency adaptation and a singular perturbation approach to construct a constant velocity traveling pulse on the real line with a general firing rate function and a general kernel. However, the existence of other phenomena are not shown. 

In this paper, we introduce a method to analyze the dynamics of a neural field model on a one- and two-dimensional domain with periodic boundary conditions and assume a smooth firing rate and an \ympree{even, periodic} kernel. Using our method, with standard numerical and analytical dynamical systems tools, we show existence and stability of traveling pulse solutions and oscillatory dynamics. \ympre{In particular, we analyze sloshing solutions on the ring and torus.} 

The neural field we consider in this paper is defined as
\begin{eqnarray}
\label{eq:u1}
\frac{\partial u(\x,t)}{\partial t} &=& -u(\x,t) + \io K(\x-\y) f(u(\y,t))\ d\y  \\
{} &+& \ve \left[q I(\x) + \io w(\x,\y)f(u(\y,t))\ d\y - g z(\x,t)\right] \nonumber, \\
\label{eq:z1}
\frac{\partial z(\x,t)}{\partial t} &=& \ve \beta [-z(\x,t)+u(\x,t)],
\end{eqnarray}
where the parameter $\varepsilon$ is small, $0 < \varepsilon \ll 1$, and $\x,\y \in \mathbb{R}^n$. For $n=2$, \ypre{the kernel function $K$ is an even function in the sense that, $K(-x,y) = K(x,-y) = K(x,y)$, and doubly periodic in the sense that $K(x+2n\pi,y+2m\pi)=K(x,y)$, for any integers $n,m$.} The function $w$ represents heterogeneity of neural connections, and $q, g, \beta$ are constants. For convenience, we will denote the domain $\Omega = [-\pi,\pi)^m$, with $m=1,2$. Thus in one-dimension the domain is a ring and in two-dimensions a torus. The variable $z(\x,t)$ represents linear adaptation \cite{pinto_ermentrout_2001_siam} and $I(\x)$ an external input to the network.  \bge{External inputs represent persistent stimuli that can be used to entrain the bump and move it to a specific location (\cite{benyishai}. We have chosen to make both the timescale of adaptation and {\em its magnitude} to be small. While there is good biological justification for the former assumption  as there are many forms of slow adaptation (\cite{johnston_wu} section 7.4), the assumption that the adaptation is small is less biological. For the existence of traveling waves, adaptation need not be small (\cite{pinto_ermentrout_2001_siam}), but in order to study how the adaptation interacts with stimuli, we need both the adaptation and the stimuli to be the same order of magnitude. The effects of large stimuli to general neural field models are not easy to analyze, so that by treating them as perturbations, we are able to consider the effects in a great deal of detail. Thus, one can regard this assumption as a starting point for the continuation of these phenomena to large amplitude stimuli and adaptation.} 

Our goal in this paper is to analyze Equations \eqref{eq:u1},\eqref{eq:z1} when $\ve$ is small. When $\ve=0$, there is a stable ``bump'' attractor, $u_0(\x)$, in the scalar neural field \eqref{eq:u1}, \ypre{i.e., a local stationary peak of $u(\x,t)$ centered at $\x=0$}.
\ympre{The bump attractor satisfies
\begin{equation*}
 u_0(\x) = \int_\Omega K(\x-\y) f(u_0(\y)) d\y,
\end{equation*}
where $u_0$ is nonconstant and even.}

Although we allow for a general \ympree{even, doubly periodic} kernel in one- and two-dimensions and a general smooth threshold nonlinearity $f$, we make particular choices for  numerical simulations. We choose $f$ as
\begin{equation*}
 f(x) = \frac{1}{\exp(-r(x-u_{th}))},
\end{equation*}
where $r=15, u_{th}=0.25$. In the one-dimensional case, we choose the kernel to be $K(x) = A + B\cos(x)$ with $A=-0.5, B=3$ unless stated otherwise. In the two-dimensional case, we form the Mexican-hat function,
\begin{equation*}
 \hat K(r) = A e^{-(r/\sigma_e)^2} - B e^{-(r/\sigma_i)^2},
\end{equation*}
where $r \equiv r(x,y,n,m) = \sqrt{(x + 2\pi n)^2 + (y + 2\pi m)^2}$. \ymp{We make the the function $\hat K$ periodic in two dimensions using the definition}
\begin{equation*}
 K(x,y) = \sum_{m=-\infty}^{\infty} \sum_{n=-\infty}^\infty \hat K(r(x,y,n,m)).
\end{equation*}
The parameters here are
\begin{equation*}
 A = \frac{1}{\sqrt{\pi} \sigma_e}, \quad B = \frac{1}{\sqrt{\pi} \sigma_i},
\end{equation*}
where $\sigma_e = 2$, and $\sigma_i = 3$. \ymp{For numerical simulations, we find it sufficient to replace the infinite sum with a finite sum from $n,m=-5$ to $n,m=5$. This is because the function $\hat K(r)$ is a decaying exponential and therefore negligible for large $r$. For example, if a bump solution remains close to the origin, contributions from terms a distance of $10\pi$ (i.e., $n$ or $m$=5) are negligible because $\exp\left(-(10\pi)^2\right) \approx 2\times10^{-429}$.

To analyze particular dynamics in more detail, we numerically compute the period kernel above, then take the Fourier truncation of this doubly periodic kernel},
\begin{equation*}
 K(x,y) = k_0 + k_1 (\cos x + \cos y) + k_2\cos x \cos y.
\end{equation*}

We now outline the organization of the paper, as follows: We reduce Equations \eqref{eq:u1},\eqref{eq:z1} to a set of integro-differential equations for the centroid of the bump solution on the ring and torus. We study bifurcations of these equations using numerical and analytical techniques to show existence and stability of constant velocity traveling bumps and sloshing bumps. Depending on the parameter values $g,q$, these traveling bumps \ympree{may traverse the domain periodically or exhibit chaos}. Next we turn to the torus domain and perform similar analyses: we study bifurcations of these equations using numerical and analytical techniques to show existence and stability of constant velocity traveling bumps. \be{In addition to the sloshing solutions found in the one-dimensional model, we also find several types of traveling bumps and modulated traveling bumps that densely fill the torus.} We also find chaotic motion in some cases. We conclude with a discussion and some contrasts to previous analyses. We remark that all figure generation code and relevant data files with documentation is available on GitHub at \url{https://github.com/youngmp/park\_and_ermentrout\_2017}

\section{Derivation of the Phase Equation}\label{sec:phase_derivation}
We start with Equations \eqref{eq:u1},\eqref{eq:z1}. Let $\tau=\ve t$ be a slow timescale and assume that both $z$ and $u$ depend only on $(\x,\tau)$.  In this case, we can integrate equation (\ref{eq:z1}) to obtain:
\[
z(\x,\tau) = z(\x,0)e^{-\beta \tau} + \beta\int_0^\tau e^{-\beta(\tau-s)} u(\x,s)\ ds.
\]
Since we are mainly interested in long term behavior, we can ignore the \ypre{first} exponentially decaying term.  With these assumptions, we obtain the following scalar integro-differential equation:
\begin{eqnarray}
\label{eq:ufull}
\ve \frac{\partial u(\x,\tau)}{\partial \tau} &=& -u(\x,\tau)  + \io K(\x-\y) f(u(\y,\tau))\ d\y  \\
{} &+&  \ve \left[q I(\x) + \io w(\x,\y)f(u(\y,\tau))\ d\y - g \beta\int_0^\tau e^{-\beta (\tau-s)}u(\x,s)\ ds\right]. \nonumber
\end{eqnarray}
We will assume $u(\x,\tau)=U(\x,\tau,\ve)$ and expand $U$ \ypre{as} a power series in $\ve$ to get an approximate solution.  Thus,
\[
U(x,\tau,\ve) = U_0(\x,\tau) + \ve U_1(\x,\tau) + O(\ve^2).
\] 
Substituting this power series into (\ref{eq:ufull}), we get (with a bit of re-arrangement):
\begin{eqnarray}
\label{eq:u0b}
0 &=& -U_0(\x,\tau) + \io K(\x-\y) f(U_0(\y,\tau)) d\y \\
\label{eq:u1b}
 (LU_1)(\x,\tau) &=& \frac{\partial U_0(\x,\tau)}{\partial \tau} - R_1(\x,\tau),
\end{eqnarray}
where
\[
(L\ypre{v})(\x,\tau) = -\ypre{v}(\x,\tau) + \io K(\x-\y)f'(U_0(\y,\tau))\ypre{v}(\y,\tau)\ d\y,
\]
and 
\[
R_1(\x,\tau) = q I(\x) + \io w(\x,\y) f(U_0(\y,\tau))\ d\y - g \beta\int_0^\tau e^{-\beta (\tau-s)}U_0(\x,s)\ ds.
\]

\begin{figure}[h!]
 \includegraphics[width=\textwidth]{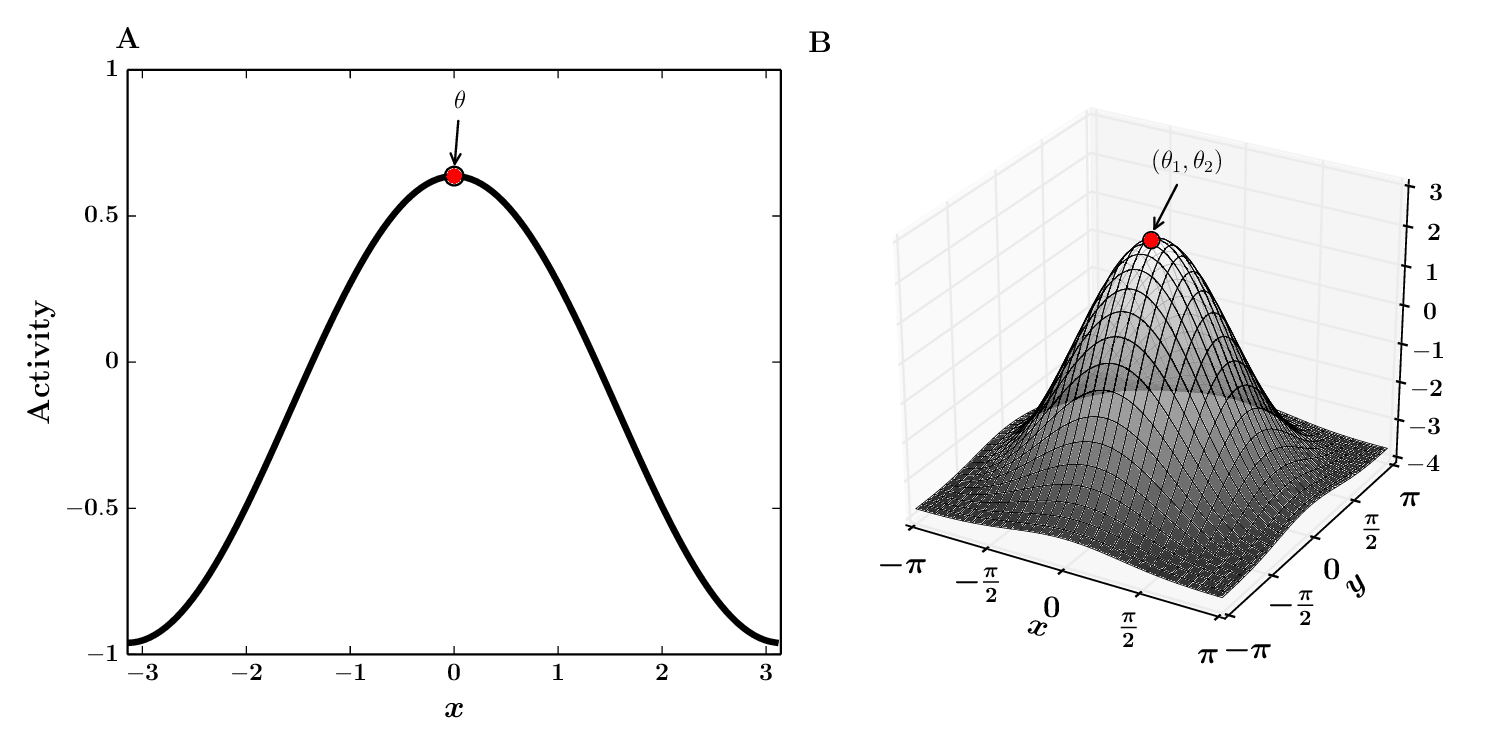}
 \caption{Numerically computed stationary bump solutions on the ring A:, and torus B:. The red circle denotes the centroid of each bump solution. On the ring, we denote the centroid by $\theta$, while we denote the centroid of the bump on the torus by $(\theta_1, \theta_2)$. Our phase model (Equation \eqref{eq:theta2}) describes shifts in the centroid.}\label{fig:ss_bumps}
\end{figure}

The equation for $U_0(\x,\tau)$ is the equation for the bump solution and since it is translation invariant, we see that
\[
U_0(\x,\tau) = u_0(\x+\q(\tau))
\] where $\q(\tau)$ is a $\tau-$dependent phase shift of the bump. Our goal, then is to determine the dynamics of $\q(\tau)$. Figure \ref{fig:ss_bumps} shows typical examples of the stationary bump $U_0(x)$ for one- and two-dimensions.

\ympree{Before continuing with the perturbation calculation, we establish a few preliminaries.} We define the compact linear operator
\[
(L_0v)(\x) = -v(\x) + \io K(\x-\y) f'(u_0(\y))v(\y)\ d\y
\] 
and establish several properties of it.  \ympre{Recall that} the bump, $u_0(\x)$ satisfies 
\[
-u_0(\x) + \io K(\x-\y)f(u_0(\y))\ d\y =0.
\] 
By making a change of variables and noting that all functions are periodic in $\x$ (that is, periodic in each of the components of $\x$), then $u_0(\x)$ satisfies
\begin{equation}
\label{eq:u0r}
-u_0(\x) + \io K(\y)f(u_0(\x-\y))\ d\y =0.
\end{equation}
Recalling that the domain is $\Omega = [-\pi,\pi)^m$, with $m=1,2$, we center $u_0$ at the origin. Thus, $u_0(\x)$ is an even periodic function of $\x$, component-wise. Let $\partial_i u(\x)$ denote the partial derivative of $u(\x)$  along the $x_i$ direction where $\x=(x_1,x_2).$    If we differentiate (\ref{eq:u0r}) along one of the axes, we see that
\[
-\pa_i u_0(\x) + \io K(\y) f'(u_0(\x-\y))\pa_i u_0(\x-\y)\ d\y =0.
\]
and changing variables again, we rewrite this as
\begin{equation}
\label{eq:nsp}
-\pa_i u_0(\x) + \io K(\x-\y) f'(u_0(\y))\pa_i u_0(\y)\ d\y =0,
\end{equation}
so we see that $L_0\pa_iu_0(\x)=0.$
In other words, the linear operator, $L_0$ has an $m-$dimensional nullspace spanned by the principle directional derivatives of $u_0(\x).$    With the natural inner product
\[
\langle u(\x),v(\x) \rangle = \io u(\x)v(\x) \ d\x 
\]
the operator $L_0$ has an adjoint
\[
(L^*v)(\x) =-v(\x) + f'(u_0(\x)) \io K(\x-\y)v(\y)\ d\y.
\] 
By multiplying equation (\ref{eq:nsp}) by $f'(u_0(\x))$, we see that the nullspace of $L^*$ is spanned by  $v^*_i(\x)=f'(u_0(\x))\pa_i u_0(\x).$ Since $u_0(\x)$ is an even periodic function, componentwise, we note that $\pa_1u_0(\x)$ is even in $x_2$ and odd in $x_1$ where $\x=(x_1,x_2)$; $v^*_1(\x)$ has the same property, while $\pa_2u_0(\x),v^*_2(\x)$ are even in $x_1$ and odd in $x_2.$  These properties imply the $\langle \pa_iu_0(\x),v^*_k(\x)\rangle=0$ when $i\ne k.$  We also have
\[
\langle \pa_i u_0(\x),v^*_i(\x) \rangle = \io f'(u_0(\x))[\pa_iu_0(\x)]^2\ d\x = \mu > 0.
\]
Finally, the Fredholm alternative holds for $L_0$. That is, for any continuous periodic function $b(\x)$,
\[
(L_0v)(\x) = b(\x)
\]   
has a bounded solution if and only if
\[
\langle v^*_i(\x), b(\x)\rangle =0
\]
for $i=1,\ldots,m$ \cite{keener}.

\begin{figure}[h!]
 \includegraphics[width=\textwidth]{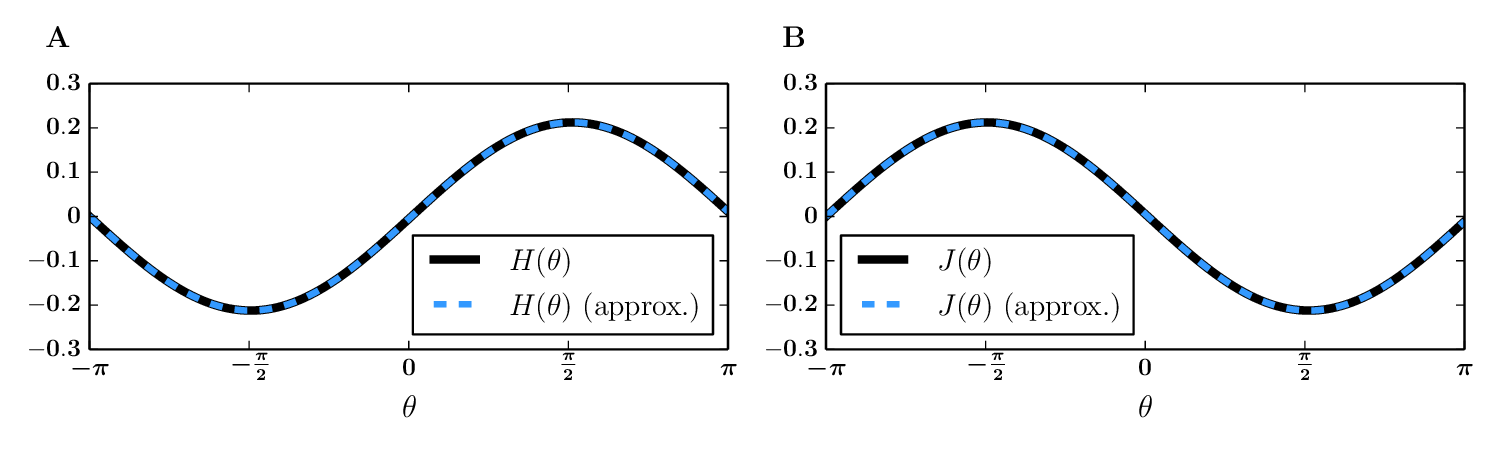}
 \caption{Numerically computed functions for the one-dimensional phase model. A: $H$ (black solid), plotted against sine-function approximation (light blue, dashed). B: $J$ (black solid), plotted against its sine-function approximation (light blue, dashed). Parameters: $I(x) = u_0(x)$ and $K(x) = A + B\cos(x)$, $A=-0.5, B=3$.}\label{fig:hj_1d}
\end{figure}

\begin{figure}[h!]
 \includegraphics[width=\textwidth]{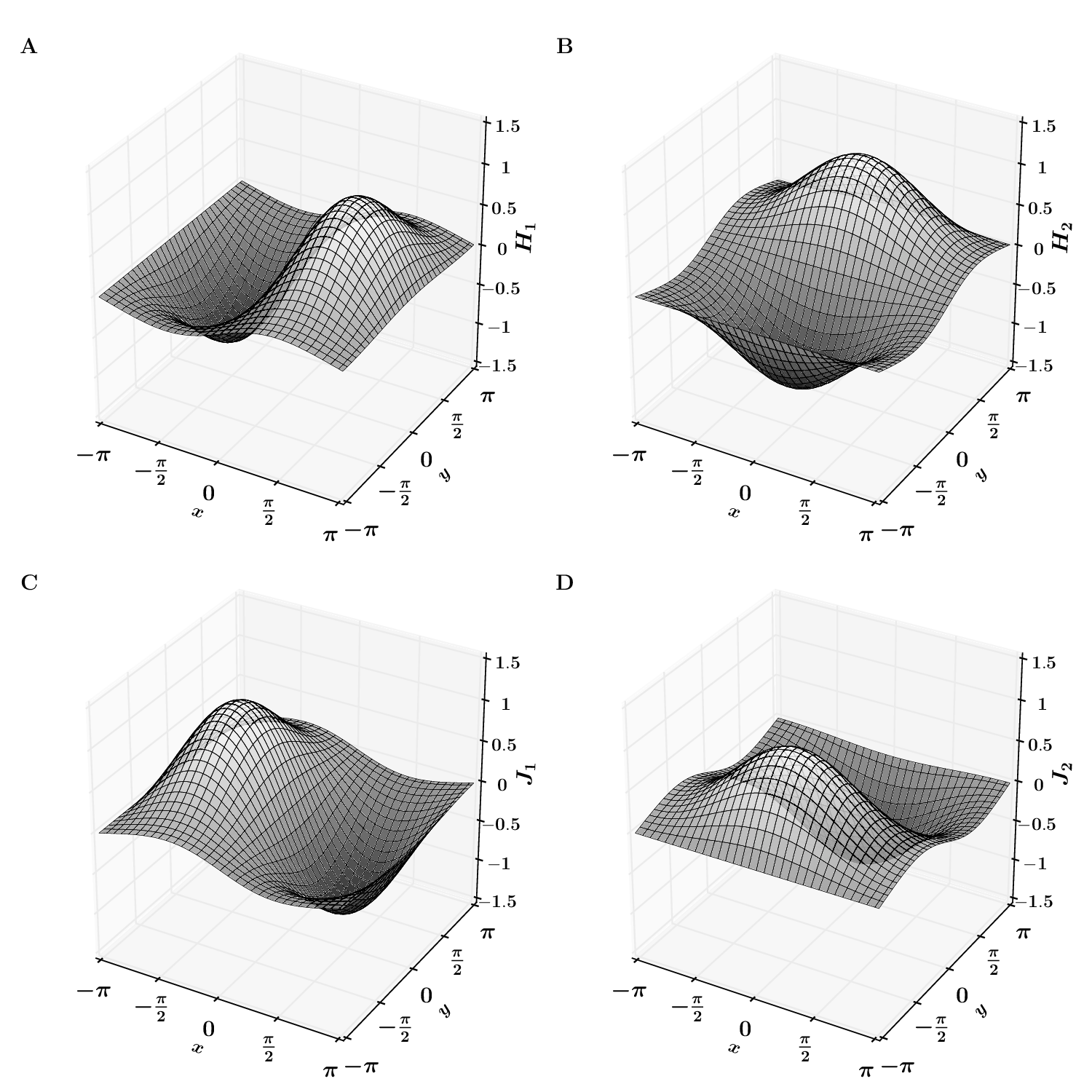}
 \caption{Numerically computed functions for the two-dimensional phase model A: $H_1$, B: $H_2$, C: $J_1$, D: $J_2$. Note that $H_2(x,y) = H_1(y,x)$. We always choose $I(x) = u_0(x)$. That is, we always use the steady-state bump as the pinning function. With this choice, $J_i = -H_i$ in 1- and 2-dimensions.}\label{fig:hj_2d}
\end{figure}

With these technical issues aside, we turn to equation (\ref{eq:u1b}), which we can rewrite as
\[
(L_0 U_1)(\x,\tau) = (\pa_1 u_0(\x+\q(\tau)),\pa_2u_0(\x+\q(\tau)))\cdot \frac{d\q(\tau)}{d\tau}   - R_1(\x,\tau)
\]
Writing $\q(\tau)=(\theta_1(\tau),\theta_2(\tau))$ and applying the $m$ conditions for the Fredholm alternative, we arrive at
\begin{equation}
\label{eq:theta}
\mu \frac{d\theta_i}{d\tau} = qJ_i(\q) + W_i(\q) -g\beta \int_0^\tau e^{-\beta(\tau-s)} H_i(\q(s)-\q(\tau))ds
\end{equation}
where
\begin{eqnarray*}
\mu &=& \io f'(u_0(\x))[\pa_i u_0(\x)]^2\ d\x,\\
J_i(\q) &=& \io f'(u_0(\x+\q))\pa_i u_0(\x+\q)I(\x)\ d\x,\\
W_i(\q) &=& \io f'(u_0(\x+\q))\pa_i u_0(\x+\q) \io w(\x,\y)f(u_0(\y))\ d\y \ d\x, \\
H_i(\q) &=& \io f'(u_0(\x))\pa_i u_0(\x) u_0(\x+\q)\ d\x.
\end{eqnarray*} 
We note that because of the symmetry of $u_0(\x)$, the functions, $H_i(\q)$ have a similar symmetry which we will exploit in the analysis of Equation \eqref{eq:theta}. The derivation here has been fairly general and holds in any dimension although we will focus only on one- and two-dimensional bumps in this model. Since both $W_i$ and $J_i$ have no explicit time dependence and act mainly as heterogeneities, we will ignore $W_i$ and focus on $J_i$ which is conveniently parameterized by $q$.  Figure \ref{fig:hj_1d} shows the functions $H(\theta), J(\theta)$ in the one-dimensional case, while Figure \ref{fig:hj_2d} shows the functions $H_i(\theta)$, $J_i(\theta)$ in the two-dimensional case.

We have reduced the problem of the bump dynamics to slow timescale phase shifts of the bump solution, represented as an integro-differential equation. For simplicity and convenience, we ignore transients by changing the limits of integration in Equation \eqref{eq:theta} from $[0,\tau]$ to $(-\infty,\tau]$, and apply the change of variables $\xi = \tau-s$, yielding
\begin{equation*}
\mu \frac{d\theta_i}{d\tau} = qJ_i(\q) -g\beta \int_0^\infty e^{-\beta \xi} H_i(\q(\tau-\xi)-\q(\tau))d\xi.
\end{equation*}
With a trivial change of notation, we arrive at the equations
\begin{equation}\label{eq:theta2}
\mu \frac{d\theta_i}{d\tau} = qJ_i(\q) -g\beta \int_0^\infty e^{-\beta s} H_i(\q(\tau-s)-\q(\tau))ds, \quad i=1,\ldots,m.
\end{equation}
We study stability properties and bifurcations in this form. Note that $H_i$ is implicitly a function of the kernel $K$.

To facilitate calculations, we first prove the following statements:
\begin{enumerate}
 \item Each function $H_i$ is odd, i.e., \ypre{$H_i(-\theta_1,-\theta_2) = -H_i(\theta_1,\theta_2)$}. In particular, $H_1$ is odd in the first coordinate and even in the second coordinate.
 \item $H_1(\theta_1,\theta_2) = H_2(\theta_2,\theta_1)$.
 \item \ymp{If the input current $I(\x)$ is defined as the steady-state bump solution, then $H_i(\q) = -J_i(\q)$.}
\end{enumerate}

\ymp{For the first statement, fix $\theta_1,\theta_2$ and consider the sum $H_1(-\theta_1,\theta_2) + H(\theta_1,\theta_2)$. By definition this sum is the sum of integrals
\begin{equation*}
 \io f'(u_0(\x))\pa_1u_0(\x)[u_0(x_1+\theta_1,x_2+\theta_2) +u_0(x_1-\theta_1,x_2+\theta_2)] d\x.
\end{equation*}
Given $x_2$, and for the sake of clarity, consider the temporary function $\Phi(x_1):=[u_0(x_1+\theta_1,x_2+\theta_2) +u_0(x_1-\theta_1,x_2+\theta_2)]$. $\Phi(x_1)$ is an even function in $x_1$ because
\begin{align*}
\Phi(-x_1) &\equiv u_0(-x_1+\theta_1,x_2+\theta_2) +u_0(-x_1-\theta_1,x_2+\theta_2)\\
 &=  u_0(x_1-\theta_1,x_2+\theta_2) +u_0(x_1+\theta_1,x_2+\theta_2)\\
 &\equiv\Phi(x_1).
\end{align*}
These lines follow by the even assumption on each coordinate oa:simplifyf the bump solution $u_0$. The remaining terms in the integrand, $\pa_1u_0(\x)$, and $f'(u_0(\x))$, are odd and even in $x_1$, respectively. Thus, the integrand is odd in $x_1$ and the integral evaluates to zero for each $x_2$ (and indeed, for each $\theta_1,\theta_2$). It follows that $H_1(-\theta_1,\theta_2) + H(\theta_1,\theta_2) = 0$, i.e., that the first coordinate is odd.

To show that the second coordinate is even, we use a similar argument. Again, fix $\theta_1,\theta_2$ and consider the sum $H_1(\theta_1,-\theta_2) - H(\theta_1,\theta_2)$. By definition, this sum is the sum of integrals
\begin{equation*}
 \io f'(u_0(\x))\pa_1u_0(\x)[u_0(x_1+\theta_1,x_2-\theta_2)-u_0(x_1+\theta_1,x_2+\theta_2)] d\x.
\end{equation*}
Given $x_1$, we redefine our temporary function $\Phi$ as $\Phi(x_2) := [u_0(x_1+\theta_1,x_2-\theta_2)-u_0(x_1+\theta_1,x_2+\theta_2)]$ and show that it is an odd function in $x_2$.
\begin{align*}
\Phi(-x_2) &\equiv u_0(x_1+\theta_1,-x_2-\theta_2) -u_0(x_1+\theta_1,-x_2+\theta_2)\\
 &=  u_0(x_1+\theta_1,x_2+\theta_2) -u_0(x_1+\theta_1,x_2-\theta_2)\\
 &= -[u_0(x_1+\theta_1,x_2-\theta_2) - u_0(x_1+\theta_1,x_2+\theta_2)]\\
 &\equiv-\Phi(x_2).
\end{align*}
Again, these lines follow by the even assmption on each coordinate of the bump solution $u_0$. The integrand term $\pa_1 u_0(\x)$ is even in the second coordinate as is the term $f'(u_0(\x))$. Thus, the integrand is odd in $x_2$ and the integral evaluates to zero for each $x_1$ (and indeed, for each $\theta_1,\theta_2$). It follows that $H_1(\theta_1,-\theta_2) - H(\theta_1,\theta_2) = 0$, i.e., that the second coordinate is even.

We have shown that $H_1$ is an odd function that is odd in the first coordinate and even in the second coordinate. The proof of $H_2$ being an odd function that is even in the first coordinate and odd in the second follows using the same arguments, or by using the second statement, which we prove next.
}

To prove the second statement, we proceed by definition.
\begin{equation*}
 H_1(\theta_1,\theta_2) = \io f'(u_0(x_1,x_2)) \pa_1 u_0(x_1, x_2) u_0(x_1+\theta_1, x_2 + \theta_2)dx_1 dx_2.
 \end{equation*}
 The steady-state bump solution is invariant under reflections about the unit line, and due to the radial symmetry of the bump solution, its partial derivatives are related by $\pa_1 u_0(x_1,x_2) = \pa_2 u_0(x_2,x_1)$. Thus,
 \begin{equation*}
 = \io f'(u_0(x_2,x_1)) \pa_2 u_0(x_2,x_1) u_0(x_1 + \theta_1, x_2 + \theta_2) dx_1 dx_2.
 \end{equation*}
Next we relabel the coordinates and flip the order of integration
\begin{equation*}
  = \io f'(u_0(x_1,x_2)) \pa_2 u_0(x_1,x_2) u_0(x_2 + \theta_1, x_1 + \theta_2) dx_1 dx_2.
\end{equation*}
Then we flip the coordinates of $u_0$, and the resulting integral is by definition $H_2(\theta_2,\theta_1)$:
\begin{align*}
 &= \io f'(u_0(x_1, x_2)) \pa_2 u_0(x_1, x_2) u_0(x_1 + \theta_2, x_2 + \theta_1) dx_1 dx_2\\
 &= H_2(\theta_2,\theta_1).
\end{align*}

To prove the third statement, suppose that a function \ympre{$\hat h$ on a periodic two-dimensional domain $[0,2\pi]\times[0,2\pi]$ is odd in the first coordinate and even in the second so that $\hat h(x_1,x_2) = \hat h(x_1,-x_2) = -\hat h(-x_1,-x_2)$. In particular, it follows that for a given value $x_2$,
\begin{align*}
 \int_0^{2\pi} \hat h(x_1,x_2) dx_1 = 0,
\end{align*}
and therefore
\begin{align*}
 \io \hat h(\x) d\x = 0.
\end{align*}
This integral property holds when $\hat h$ is even in the first coordinate and odd in the second with a similar argument.}

If we choose $I(\x)$ to be the steady-state bump, then
\begin{align*}
 J_i(\q) &= \io f'(u_0(\x+\q))\pa_i u_0(\x+\q)u_0(\x)\ d\x\\
 &= \io f'(u_0(\x))\pa_i u_0(\x)u_0(\x-\q)\ d\x.
\end{align*}
Then taking the sum $H_i(\q) + J_i(\q)$ yields
\begin{equation*}
 H_i(\q) + J_i(\q) = \io f'(u_0(\x))\pa_i u_0(\x)[u_0(\x+\q) + u_0(\x-\q)]\ d\x.
\end{equation*}
\ympre{For a given $\q$, the term $[u_0(\x+\q) + u_0(\x-\q)]$ in the integrand is even in both coordinates. The remaining term, $f'(u_0(\x))\pa_i u_0(\x)$, when $i=1$ ($i=2$), is odd (even) in the first coordinate and even (odd) in the second. Therefore, when $i=1$ ($i=2$), the integrand is an odd function in the first (second) coordinate and the integral evaluates to zero. It follows trivially that}
\begin{equation}\label{eq:hi_to_ji}
  H_i(\q) = - J_i(\q).
\end{equation}
This property remains true on the ring using the same argument.

These statements will come in useful in the sections to follow. We now proceed with an analysis of the reduced equations on the ring domain.

\be{\bf{Remark} We have assumed {\em linear} adaptation in our derivation of the reduced model, but, this is not necessary. We could replace Equation (\ref{eq:z1}) by 
\[
 \frac{\partial z(\x,t)}{\partial t} = \ve \beta [-z(\x,t)+M(u(\x,t))]
\]
where $M(u)$ is an arbitrary monotonically increasing continuously differentiable function. In this case, we find
\[
H_i(\q) = \io f'(u_0(\x))\pa_i u_0(\x) M(u_0(\x+\q))\ d\x.
\]
The new version of $H_i$ has exactly the same properties as the linear case since $M(u_0(x))$ is an even function and its derivative with respect to $x$ is an odd function. }

\section{The Ring Domain}\label{sec:1d}
\ypre{In this section, we choose the domain $\Omega$ to be the ring. First, we thoroughly analyze the full neural field model through a bifurcation analysis. We then turn to Equation \eqref{eq:theta2} on the ring and perform the same bifurcation analysis and through analytical study.}

\subsection{Equivalent Neural Field Model on the Ring}
To classify the bifurcations of the full neural field model on the ring, we \ymp{transform} the equations to an equivalent 6-dimensional system of ODEs, \ymp{allowing} us to use dynamical systems software and techniques to analyze the model. \ymp{Recall that for numerical simulations on the ring, we choose a cosine kernel $K(x) = A + B\cos(x)$. This technique and choice of kernel is the same as that used in \cite{laing_noise-induced_2001}, where as part of the study they analyze a rate model similar to the model in the current study, but in contrast, the adaptation and input current terms are input directly to the firing rate function. They provide sufficient detail with regards to transforming their rate model to a system of ODEs, but as the details differ from our model, we include the derivation of our model here (in particular they include a phase lag between the peak of the bump activity $u$ and the peak of the adaptation activity $z$ which results in slightly different equations).

Note that with this choice of kernel, the bump solution is also sinusoidal and without loss of generality takes the form $u_0(x) = C + D\cos(x)$.  For simplicity we choose $J(x) = u_0(x)$. We are now ready to transform the equations.

Since the functions $u(x,t),z(x,t)$ are periodic in $x$, we expand them in a Fourier series,
\begin{align*}
 u(x,t) &= \hat a_0(t) + \sum_{n=1}^\infty \hat a_n(t) \cos n x + \hat b_n(t) \sin n x,\\
 z(x,t) &= \hat c_0(t) + \sum_{n=1}^\infty \hat c_n(t) \cos n x + \hat d_n(t) \sin n x.
\end{align*}
and plug into equations \eqref{eq:u1},\eqref{eq:z1}. First, a direct substitution into the dynamics of $u$ yields
\begin{align*}
 \hat a_0' +& \sum_{n=1}^\infty \hat a_n' \cos(nx) + \hat b_n \sin(nx)\\
 =& -\hat a_0 -\left[ \sum_{n=1}^\infty \hat a_n \cos nx + \hat b_n \sin nx \right]\\
 &+ A \io f(u(y,t)) dy\\
 &+ B\cos(x)\io \cos(y)f(u(y,t)) dy\\
 &+ B\sin(x)\io \sin(y)f(u(y,t)) dy\\
 &+ \ve \left[q (C+D\cos(x)) - g\left(\hat c_0(t) + \sum_{n=1}^\infty \hat c_n(t) \cos n x + \hat d_n(t) \sin n x\right) \right].
\end{align*}
We have used the elementary trigonometric identity $\cos(x-y) = \cos(x)\cos(y) + \sin(x)\sin(y)$ to separate the kernel into multiple integrals. A direct substitution into the dynamics of $z$ yields
\begin{align*}
 \hat a_0' +& \sum_{n=1}^\infty \hat a_n' \cos(nx) + \hat b_n \sin(nx)\\
 =& \ve \beta\left[ - \hat c_0(t) - \sum_{n=1}^\infty \hat c_n(t) \cos n x + \hat d_n(t) \sin n x \right.\\
 &\quad \left.+ \hat a_0(t) + \sum_{n=1}^\infty \hat a_n(t) \cos n x + \hat b_n(t) \sin n x\right]
\end{align*}
Next, we group like terms in the Fourier basis, starting with the Fourier coefficients of $u$:
\begin{align*}
 \hat a_0' &= -\hat a_0 + A \io f(u(y,t)) dy + \ve [q C - g\hat c_0]\\
 \hat a_1' &= -\hat a_1 + B\cos(x)\io \cos(y)f(u(y,t)) dy + \ve [q D -g \hat c_1]\\
 \hat a_2' &= -\hat a_2 + \ve [-g \hat c_2]\\
 \hat a_3' &= -\hat a_3 + \ve [-g \hat c_3]\\
 &\vdots\\
\end{align*}
and
\begin{align*}
  \hat b_1' &= -\hat b_1 + B\sin(x)\io \sin(y)f(u(y,t)) dy + \ve [-g \hat d_1]\\
  \hat b_2' &= -\hat b_2 + \ve [-g \hat d_2]\\
  \hat b_3' &= -\hat b_3 + \ve [-g \hat d_3]\\
  &\vdots\\
\end{align*}
We repeat this grouping for the Fourier coefficients of $z$:
\begin{align*}
 \hat c_0 &= \ve \beta(-\hat c_0 + \hat a_0)\\
 \hat c_1 &= \ve \beta(-\hat c_1 + \hat a_1)\\
 &\vdots
\end{align*}
and
\begin{align*}
 \hat d_0 &= \ve \beta(-\hat d_0 + \hat b_0)\\
 \hat d_1 &= \ve \beta(-\hat d_1 + \hat b_1)\\
 &\vdots
\end{align*}
The pattern is clear at this point: The coefficients of all Fourier modes greater than 1 satisfy
\begin{align*}
a_i' &= -a_i - \ve  g b_i,\\
b_i' &= \ve \beta (-b_i +a_i),
\end{align*}
where $a_i$ are placeholders for the Fourier coefficients of $u$ and $b_i$ are placeholders for the Fourier coefficients of $z$. Through an elementary stability analysis, all solutions to these equations decay to zero so they are unnecessary to consider. We proceed with the remaining nontrivial terms,
\begin{align*}
 u(x,t) &= a_0(t) + a_1(t) \cos x + a_2(t) \sin x,\\
 z(x,t) &= b_0(t) + b_1(t) \cos x + b_2(t) \sin x.
\end{align*}
Note that these are still the first two Fourier modes, but we have dropped the hat notation and relabeled the coefficients. } Using this notation, we have the system 
\begin{align*}
 a_0' &= -a_0 + A\int_\Omega f(u(y,t))dy + \ve (q\ymp{C} - g b_0),\\
 a_1' &= -a_1 + B\int_\Omega \cos(y) f(u(y,t)) dy + \ve ( q \ymp{D} - g b_1),\\
 a_2' &= -a_2 + B\int_\Omega \sin(y) f(u(y,t)) dy - \ve g b_2,\\
 b_i' &= \ve \beta (-b_i + a_i), \quad i=0,\ldots 2.
\end{align*}
Note that we do not need to explicitly write the full Fourier series of $f$ or extract any of its coefficients. In fact, the Fourier modes of $f(u)$ greater than 1 vanish as we will now show.

\ymp{
Consider the Fourier series of $f(u(x,t))$:
\begin{align*}
 f(u(x,t)) = \hat \alpha_0(t)  + \sum_{n=1}^\infty \hat \alpha_n(t) \cos(nx) + \hat\beta_n(t) \sin(nx)
\end{align*}
This expansion exists because $f$ is bounded and integrable on $[0,2\pi]$. We now evaluate each integral, $\int_\Omega f(u(y,t))dy$, $\int_\Omega \cos(y) f(u(y,t)) dy$, and $\int_\Omega \sin(y) f(u(y,t)) dy$ in turn. First,
\begin{align*}
 \int_\Omega f(u(y,t))dy &= \io \alpha_0 + \sum_{n=1}^\infty \alpha_n \cos(ny) + \beta_n \sin(ny) dy\\
 &= \alpha_0  \io dy + \sum_{n=1}^\infty \alpha_n \io \cos(ny)dy + \beta_n \io \sin(ny) dy\\
 &= 2\pi\alpha_0.
\end{align*}
Next,
\begin{align*}
 &\int_\Omega \cos(y) f(u(y,t)) dy = \cos x\io \cos(y) \left[ \alpha_0  + \sum_{n=1}^\infty \alpha_n \cos(ny) + \beta_n \sin(ny) \right]dy\\
 &=\alpha_0 \cos x \io \cos(y)dy + \sum_{n=1}^\infty \alpha_n \io \cos(y)\cos(ny)dy + \beta_n\io \cos(y)\sin(ny) dy\\
 &= \pi \alpha_1
\end{align*}
and finally,
\begin{align*}
 &\int_\Omega \sin(y) f(u(y,t)) dy = \sin x\io \sin(y) \left[ \alpha_0  + \sum_{n=1}^\infty \alpha_n \cos(ny) + \beta_n \sin(ny) \right]dy\\
 &= \alpha_0 \sin x \io \sin(y)dy + \sum_{n=1}^\infty \alpha_n \io \sin(y)\cos(ny)dy + \beta_n\io \sin(y)\sin(ny) dy\\
 &= \pi \beta_1.
\end{align*}
Thus, the nonlinearity $f$ in the integrand only appears in the dynamics of the first few Fourier coefficients. At each time step in the numerics, we compute the integrals $\int_\Omega f(u(y,t))dy$, $\int_\Omega \cos(y) f(u(y,t)) dy$, and $\int_\Omega \sin(y) f(u(y,t)) dy$ using Riemann integration at each time step as it is more straightforward than extracting the necessary Fourier coefficients.

}

\begin{figure}[h!]
 \includegraphics[width=\textwidth]{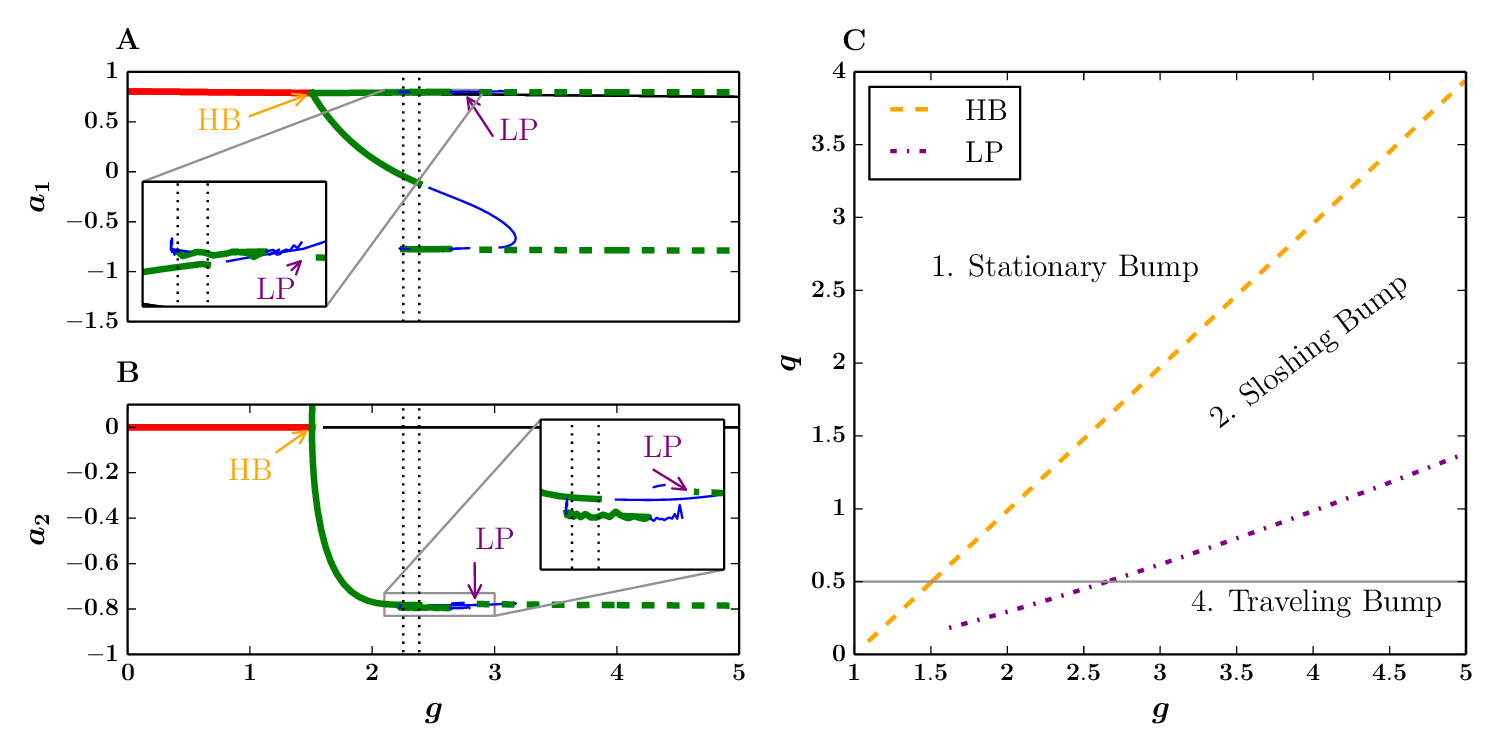}
 \caption{1- and 2-parameter bifurcation diagrams of the neural field model on the ring. Solid red lines: Stable equilibrium. Solid Green: Stable periodic solutions. Solid blue: Unstable periodic solutions. Solid black: unstable equilibrium. (A,B) Bifurcation diagram in $a_1$ and $a_2$ for fixed $q=0.5$. As $g$ increases from $0$ to $5$, the system undergoes a Hopf bifurcation (HB, orange). Solutions here slosh with a small deviation from the origin. By increasing $g$, we see a region of bistability (shown in the insets with the interval of bistability marked by vertical dotted black lines), marking the emergence of large-sloshing solutions alongside sloshing solutions. Next, the system reaches a limit point (LP, purple) beyond which there exists a traveling bump solution. For panel B:, the branches of the Hopf bifurcation are symmetric over the $x$-axis, thus we only show one branch. C: Two parameter bifurcation diagram in $g$ and $q$. To the left of the Hopf bifurcation (HB, dashed orange line), there is only a stationary bump solution (1.). Motion exists to the right of this dividing line in the form of sloshes (2.) and a traveling bump solution (4.). }\label{fig:full_bifurcations}
\end{figure}

We focus our numerical studies on the coefficients $a_1$ and $a_2$ because they produce the most salient features of the bump solution (the $a_0$ coefficient changes as a function of time, but only up to order $O(\varepsilon)$, while the $b_i$ terms represent aggregate behavior of the adaptation variable $z$). By following the fixed points and oscillatory behavior in $a_1$ and $a_2$, we produce a bifurcation diagram of this system in Figure \ref{fig:full_bifurcations}.

Figure \ref{fig:full_bifurcations} shows that there are three main solution types: the pinned or stationary bump, the sloshing bump, and the traveling bump, which traverses the ring at some finite speed. In addition there are small regions of bistability between the sloshing bump and the traveling bump.

\subsection{Phase Model on the Ring}
We now turn to the analysis of the phase dynamics on the ring. The analysis to follow depends on proving the following statements:
\begin{enumerate}
 \item \ypre{$H(0) = 0$},
 \item $H'(0) > 0$,
 \item If $K(x) =  A + B\cos(x)$, then $H(\theta) = A'\sin(\theta)$, $A'>0$, where $A'$ depends on the parameters $A,B$.
\end{enumerate}
Recall the \ypre{scalar} version of the functions $H_i(\q)$ and $v_i^*(\x)$:
\begin{align*}
 H(\theta) &= \int_\Omega u^*(y) u_0(y + \theta) dy\\
 u^*(x) &= f'(u_0(x))u_0'(x),
\end{align*}
Because $u_0$ is even, it follows that $f'(u_0)$ is even, $u_0'$ is odd, and therefore $u^*$ is odd. Noting that
\begin{equation*}
 H(0) = \int_\Omega u^*(y) u_0(y)dy,
\end{equation*}
where the function $u^*(y) u_0(y)$ is odd, the first statement follows.

For the second statement, we follow the definitions to arrive at
\begin{equation*}
 H'(0) = \int_\Omega f'(u_0(y))u_0'(y)u_0'(y)dy.
\end{equation*}
The function $f$ is an increasing sigmoidal, thus $f'>0$. In addition, $u_0'^2 >0 $. Thus, $H'(0) > 0$.

Next, we prove the third statement. With the kernel choice $K(x) = A + B\cos(x)$, the steady-state bump solution is some shifted multiple of cosine, $u_0(x) = C + D\cos(x)$, where $C,D$ implicitly depend on the kernel parameters $A,B$. Plugging this $u_0$ into $H(\theta)$ yields
\begin{equation*}
 H(\theta) = \int_{-\pi}^{\pi} f'(C+D\cos(y))[-D\sin(y)][C+D\cos(y+\theta)]dy.
\end{equation*}
Let $h(y) = f'(C+D\cos(y))[D\sin(y)]$, which is an odd function. Recalling that $\cos(y+x) = \cos(y)\cos(x) - \sin(y)\sin(x)$, $H$ simplifies to
\begin{equation*}
\begin{split}
 H(\theta) &= -C\int_{-\pi}^\pi h(y) dy - \int_{-\pi}^{\pi} h(y)D[\cos(y)\cos(\theta) - \sin(y)\sin(\theta)]dy.\\
\end{split}
\end{equation*}
Because $h(y)$ is odd, some integrals cancel, and we are left with
\begin{equation*}
 H(\theta) = A'\sin(\theta),
\end{equation*}
where $A' = D\int_{-\pi}^{\pi} h(y) \sin(y)dy$.

From these statements, it follows that $\left.\frac{dJ}{d\theta}\right |_{\theta=\overline\theta}<0$ and $J(\overline\theta)=0$ where $\overline\theta$ represents a steady-state bump peak. WLOG, we let $\overline \theta = 0$ because we generally choose the center of the steady-state bump to be the origin. 

\be{\bf{Remark:} For the more general adaptation (c.f. above), as long as $M(u)$ is differentiable and monotonically increasing, we still have that $H'(0)>0.$ } 

\subsubsection{Equivalent Phase Model on the Ring}
We next show that there are really only two relevant parameters. We can rescale time to obtain
\begin{equation*}
 \mu \beta \frac{d\theta}{d\tau} = qJ(\theta) - g\int_0^\infty e^{-s} H(\theta(\tau-s)-\theta(\tau)) ds,
\end{equation*}
where we have re-used $\tau$ as the now scaled time $\beta\tau$. Next, divide by $\mu \beta$ to obtain
\begin{equation}\label{eq:phs_rscl}
 \frac{d\theta}{d\tau} = \hat q J(\theta) - \hat g \int_0^\infty e^{-s} H(\theta(\tau-s)-\theta(\tau)) ds,
\end{equation}
with $\hat q = \frac{q}{\mu\beta}$ and $\hat g = \frac{g}{\mu \beta}$. \ympree{This rearrangement shows that making adaptation slower by decreasing $\beta$ is equivalent to increasing the rescaled parameters $\hat g$ and $\hat q$}. For analytic calculations, we will often reference this equation without the hats on the parameters.

For numerical studies of bifurcations in this system, we let $H(\theta) = A'\sin\theta$, from which $J$ follows immediately (see Equation \eqref{eq:hi_to_ji} and statement 2 above). We once more abuse notation and absorb $A'$ into $\hat g$ and into $\hat q$, then drop the hats. So we will now study
\begin{equation*}
 \frac{d\theta}{d\tau} = -q \sin(\theta) - g \int_0^\infty e^{-s} \sin(\theta(\tau-s)-\theta(\tau)) ds,
\end{equation*}

To numerically integrate this phase equation, we rewrite this differential equation as a system of three equations by exploiting basic differentiation properties of integrals. To begin, we use a trigonometric identity to rewrite the integral in the right-hand side
\begin{equation*}
\begin{split}
 \frac{d\theta}{d\tau} &= -q \sin(\theta) - g \int_0^\infty e^{-s} \sin(\theta(\tau-s)-\theta(\tau)) ds\\
 &= - q\sin(\theta) - g[\cos(\theta) S(\tau) - \sin(\theta) C(\tau)],
\end{split}
\end{equation*}
where
\begin{align*}
 S(\tau) &= \int_0^\infty e^{-s} \sin(\theta(\tau-s)) ds,\\
 C(\tau) &= \int_0^\infty e^{-s} \cos(\theta(\tau-s)) ds.
\end{align*}
With the change of variables $s'=\tau-s$, $S,C$ become
\begin{equation*}
\begin{split}
 S(\tau) &= \int_{-\infty}^\tau e^{-(\tau-s')} \sin(\theta(s')) ds',\\
 C(\tau) &= \int_{-\infty}^\tau e^{-(\tau-s')} \cos(\theta(s')) ds'.
\end{split}
\end{equation*}
By differentiating, we rewrite $S$ and $C$ as ODEs:
\begin{equation*}
\begin{split}
 \frac{dS}{d\tau} &= -S(\tau) + \sin\theta,\\
 \frac{dC}{d\tau} &= -C(\tau) + \cos\theta.
\end{split}
\end{equation*}

We have transformed a single integro-differential equation into a system of three ODEs, simplifying the numerics considerably:
\begin{equation*}
\begin{split}
 \frac{d\theta}{d\tau} &= - q\sin(\theta) - g[\cos(\theta) S(\tau) - \sin(\theta) C(\tau)]\\
 \frac{dS}{d\tau} &= -S(\tau) + \sin\theta,\\
 \frac{dC}{d\tau} &= -C(\tau) + \cos\theta.
\end{split}
\end{equation*}

\begin{figure}[h!]
 \includegraphics[width=\textwidth]{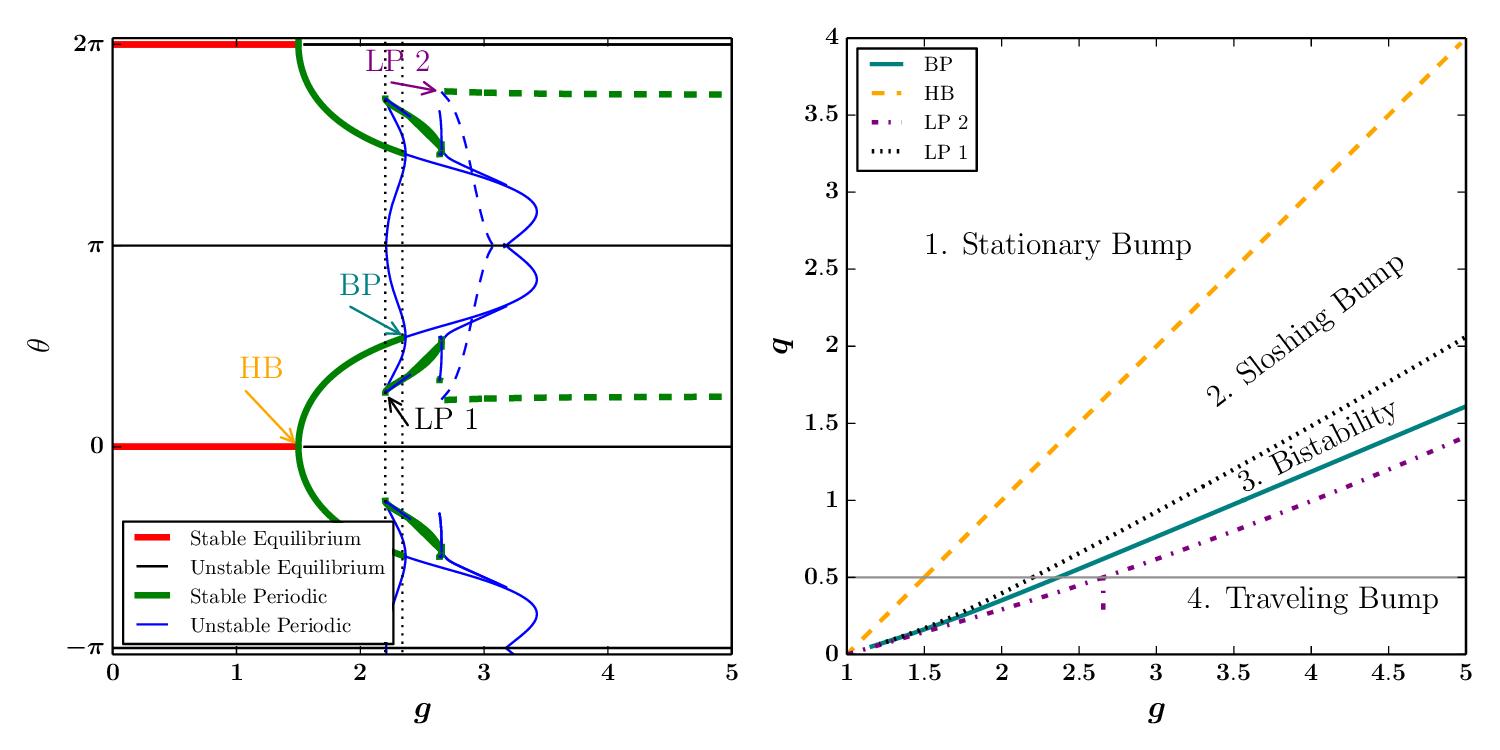}
 \caption{1- and 2-parameter bifurcation diagrams of the phase equation on the ring. (A,B) Bifurcation diagram in for fixed $q=0.5$. As $g$ increases from $0$ to $5$, the system undergoes a Hopf bifurcation (HB, orange) then produces a limit point (LP 1, black), a branch point (BP, teal), and another limit point (LP 2, purple), respectively. Between the limit point LP1 and branch point BP, there is bistability, the interval of which is denoted by vertical dotted black lines. Beyond the second limit point LP2, there exists a traveling bump solution. This traveling bump solution is distinct from the equilibria and periodic solutions denoted by solid lines, thus we label it with a dashed green line. B: Two parameter bifurcation diagram in $g$ and $q$. To the left of the Hopf bifurcation (HB, dashed orange line), there is only a stationary bump solution (1.). Motion exists to the right of this dividing line in the form of sloshes (2., 3.) and a traveling bump solution (4.). }\label{fig:phase_bifurcations}
\end{figure}

The bifurcation diagram in Figure \ref{fig:phase_bifurcations} summarizes the dynamics of the phase model on the ring. On the left panel, we fix a parameter value $q=0.5$ and as we vary the parameter $g$, the system transitions from steady-state to sloshing solutions, then to a co-existence of large-amplitude and relatively small amplitude sloshing solutions, and eventually to a steady traveling pulse. On the right panel, we find that the parameter space is separated into several regions. In particular, for $q \geq 0$ arbitrarily small, there exists a traveling bump for some nonzero $g$.

In the following sections, we analyze the existence of these bifurcations including the Hopf bifurcation leading to sloshing solutions, and the saddle-node bifurcation leads to the constant-velocity traveling bump.

\subsection{Constant Velocity Bump Solution on the Ring}\label{subsec:pitchfork}
To show the existence of a constant velocity traveling bump solution, we require that $q=0$ and $g > 0$. For the first part of this analysis, we do not require the kernel to take a particular form. We only require the kernel to be even and admit a steady-state bump solution to Equation \eqref{eq:u1}. We make a traveling bump ansatz, $\theta(\tau) = \nu\tau$, where $\nu$ corresponds to the traveling bump velocity. We first determine the existence and stability of the zero velocity bump solution. Plugging the ansatz into Equation \eqref{eq:phs_rscl} yields
\begin{equation}\label{eq:nu_g}
\begin{split}
 \nu &= -g \int_0^\infty e^{-s}H(-\nu s)ds\\
 &= g \int_0^\infty e^{- s}H(\nu s)ds,
\end{split}
\end{equation}
where the last line follows by the oddness of $H$. Because $H(0)=0$, $\nu=0$ is a solution. To determine the stability of the zero velocity solution, we consider a small perturbation, $\theta(\tau) = \nu \tau + \ve \psi$. By plugging this perturbation into Equation \eqref{eq:phs_rscl}, we extract the dynamics of the perturbed variable $\psi$,
\begin{equation}\label{eq:psi}
 \frac{d\psi}{d\tau} = -g \int_0^\infty e^{-s} H'(\nu s)[\psi(\tau-s)-\psi(\tau)]ds.
\end{equation}
Assuming $\psi(\tau) = e^{\lambda \tau}$ and $\nu=0$, we obtain the stability equation,
\begin{equation*}
 \lambda = -g \int_0^\infty e^{-s}H'(0) \left[e^{-\lambda s}-1\right]ds.
\end{equation*}
We integrate the right-hand side and rearrange to yield
\begin{equation*}
 \lambda = gH'(0) \frac{\lambda}{1+\lambda}.
\end{equation*}
Thus, either $\lambda = 0$, or $\lambda = -1 + gH'(0)$. Moreover, the zero velocity solution becomes unstable when $g > 1/H'(0)$.

\begin{figure}[h!]
 \includegraphics[width=\textwidth]{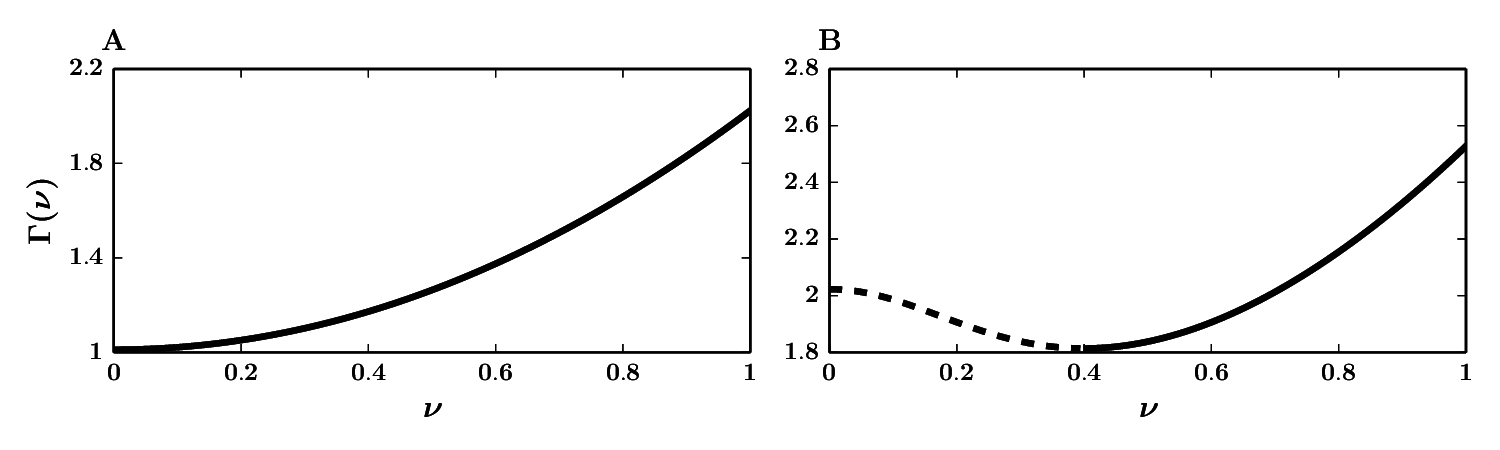}
  \caption{Examples of the function \ypre{$\Gamma(\nu)$}. A: \ypre{$\Gamma$} constructed using our usual $H$ function, $H(x) = \sin(x)$. B: \ypre{$\Gamma$} constructed using a different $H$ function, $H(x) = \sin(x) - 0.25\sin(2 x)$, resulting from a different choice of kernel. The dashed black line represents an unstable traveling bump velocity.}\label{fig:g_nu}
\end{figure}

In general, we may view the relationship between $g$ and $\nu$ by rearranging Equation \eqref{eq:nu_g} into the function
\begin{equation}\label{eq:g_nu}
 \ypre{g = \Gamma}(\nu) := \frac{\nu}{\int_0^\infty e^{-s}H(\nu s)ds}.
\end{equation}
We show examples of \ypre{$\Gamma$} in Figure \ref{fig:g_nu}. In the left panel, the relationship between the adaptation strength $g$ and bump velocity $\nu$ is straightforward: as $g$ increases, there is some critical value $\nu$ where a nonzero velocity traveling bump exists. However, the choice of kernel may change the shape of \ypre{$\Gamma$}, and therefore change the relationship between $g$ and $\nu$, as well as the stability of traveling bump solutions. For example, a kernel of the form $K(x) = a + b\cos(x) + c\cos(2x)$ results in an $H$ function of the form
\begin{equation*}
 H(\theta) = a'\sin(\theta) + b'\sin(2\theta). 
\end{equation*}
Using this $H$ function to plot \ypre{$\Gamma$} results in the right panel of Figure \ref{fig:g_nu}. The branch with negative slope represents \ypre{another} traveling bump solution. We now show that if $\ypre{\Gamma}'(\nu)<0$, then the traveling bump with velocity $\nu$ is unstable.

Recall again Equation \eqref{eq:psi}. Assuming  $\psi(\tau) = e^{\lambda \tau}$ and $\nu \neq 0$, we obtain the stability equation,
\begin{equation}\label{eq:wave_stability}
 f(\lambda) \equiv 1 + g\int_0^\infty e^{-s}H'(\nu s) \left[\frac{e^{-\lambda s}-1}{\lambda}\right]ds.
\end{equation}
To prove the statement, we seek to show that $\lim_{\lambda \rightarrow \infty} f(\lambda)>0$ and $\lim_{\lambda \rightarrow 0} f(\lambda) < 0$. Then by continuity of $f$, there exists a positive root to Equation \eqref{eq:wave_stability}. We take each limit in turn, \ypre{starting with the limit as $\lambda \rightarrow \infty$}.

\begin{align*}
\lim_{\lambda \rightarrow \infty} f(\lambda) &= 1 + g\int_0^\infty e^{-s} H'(\nu s) \lim_{\lambda \rightarrow \infty} \left[ \frac{e^{-\lambda s} - 1}{\lambda} \right]ds\\
&= 1 > 0.
\end{align*}
\ypre{Thus, the positive $\lambda$ limit is positive}. For the other limit, we rearrange Equation \eqref{eq:g_nu} into
\begin{equation*}
 \ypre{\Gamma}(\nu)\ypre{D}(\nu) = \nu,
\end{equation*}
where $\ypre{D}(\nu) = \int_0^\infty e^{-\beta s}H(\nu s)ds$, and differentiate with respect to $\nu$ to obtain
\begin{equation*}
 \ypre{\Gamma}(\nu)\ypre{D}'(\nu) + \ypre{\Gamma}'(\nu) \ypre{D}(\nu) = 1.
\end{equation*}
Solving for $\ypre{\Gamma}'(\nu)$ yields
\begin{equation*}
 \ypre{\Gamma}'(\nu)  = \frac{1-\ypre{\Gamma}(\nu)\ypre{D}'(\nu)}{\ypre{D}(\nu)}.
\end{equation*}
Note that $\ypre{D(\nu)} > 0$ at least within a neighborhood of $\nu=0$ since $H(0)=0$ and $H'(0)>0$. In addition, $\ypre{D'(\nu)} = \int_0^\infty e^{-s}H'(\nu s) sds$. Using the hypothesis that $\ypre{\Gamma'(\nu)} < 0$, we have the inequality
\begin{equation*}
 1 < \ypre{\Gamma(\nu)} \ypre{D'(\nu)}.
\end{equation*}
We use this fact in the next limit
\begin{align*}
\lim_{\lambda \rightarrow 0} f(\lambda) &= 1 + g\int_0^\infty e^{-s} H'(\nu s) \lim_{\lambda \rightarrow 0} \left[ \frac{e^{-\lambda s} - 1}{\lambda} \right]ds\\
&= 1 - g\int_0^\infty e^{-s} H'(\nu s) s ds\\
&= 1 - g\ypre{D'(\nu)}\\
&= 1 - \ypre{\Gamma(\nu)D'(\nu)} < 0.
\end{align*}
\ypre{Thus, the zero $\lambda$ limit is negative. Because $f(0)$ is negative, and $f(\lambda)$ is positive for asymptotically large values of $\lambda$, there exists a positive root $\lambda$ of $f(\lambda)$ by continuity. It follows that branches of $\ypre{\Gamma(\nu)}$ with negative slope indicate an unstable traveling bump at least within a neighborhood of $\nu=0$.}

\begin{figure}[h!]
 \includegraphics[width=\textwidth]{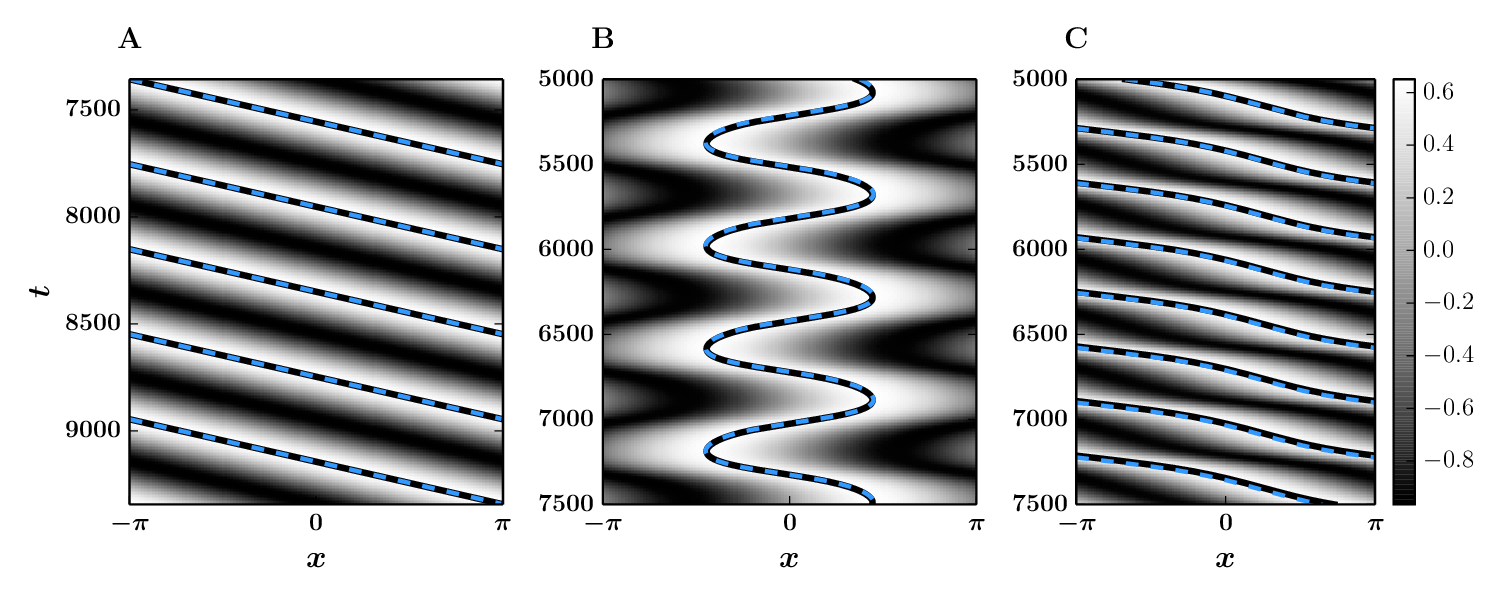}
 \caption{Dynamics of the traveling bump on the ring. Each row represents the bump solution at a particular time. White represents high activity, while black represents low or inhibited activity. The numerical centroid (black solid) is plotted against the analytic prediction (dashed blue). A: A constant-velocity bump, $g=3.5$,$q=0$. B: A \ympree{sloshing} bump, $g=3$, $q=1$. C: A non-constant velocity bump, $g=5.5$, $q=1$. For each panel, we shift the theory along the time axis to show qualitative agreement with the numerics. Parameter $\ve = 0.01$.}\label{fig:1d_dynamics}
\end{figure}

For the next part of this analysis, we show how to compute a formula for the velocity of the traveling bump when the kernel is $K(x) = A + B\cos(x)$. With this kernel, the $H$ function is proportional to $\sin(x)$, and Equation \eqref{eq:nu_g} becomes explicitly computable. Computing the integral results in a formula for the nontrivial bump velocity $\nu$,
\begin{equation}\label{eq:bump_speed}
 \nu = \pm \sqrt{g- 1}.
\end{equation}
Equation \eqref{eq:bump_speed} corresponds to the branches of a pitchfork bifurcation in the velocity of the traveling bump. We show a particular example of a constant-velocity traveling bump in Figure \ref{fig:1d_dynamics}A. \ympree{We note that any odd $H$ will lead to a pitchfork bifurcation to a traveling bump.  
\be{In particular, it is trivial to derive the following bifurcation equation:}
\begin{equation*}
 \nu^2 = (1-g H'(0))/g H'''(0).
\end{equation*}
This equation tells us that the pitchfork bifurcation is super-critical if $H'''(0) < 0$ and sub-critical otherwise.}

\subsection{Andronov-Hopf Bifurcation on the Ring}\label{subsec:hopf}
We now prove the existence of a Hopf bifurcation. For this analysis, we do not require $H$ or $J$ to take a particular form. However, we do require that $H$ and $J$ be sufficiently differentiable, along with the properties $H(0)=J(0)=0$, $H'(0)>0$,$J'(0) < 0$, $H$ odd, and $g,q>0$.

Consider again the simplified phase model, Equation \eqref{eq:phs_rscl}. Let us fix $q$ and absorb the parameter into $J$. We write $J$ and $H$ as Taylor expansions,
\begin{align*}
 J(\theta) &= j_1\theta + j_2 \theta^2 + j_3 \theta^3,\\
 H(\theta) &= h_1\theta + h_3 \theta^3.
\end{align*}
Then to first order,
\begin{equation*}
\frac{d\theta}{d\tau} = j_1\theta - gh_1 \int_0^\infty e^{-s} [\theta(\tau-s)-\theta(\tau)] ds.
\end{equation*}
Letting $\theta = e^{\lambda t}$ and rearranging the resulting equation yields
\begin{equation*}
 \lambda = j_1 + gh_1 - \frac{gh_1}{\lambda + 1},
\end{equation*}
or equivalently,
\begin{equation*}
 \lambda^2 + \lambda(1-j_1 - g h_1) - j_1 = 0.
\end{equation*}
Since $j_1 < 0$ and $h_1 > 0$, there exists a Hopf bifurcation when
\begin{equation*}
 g^* = \frac{1-j_1}{h_1}.
\end{equation*}

This bifurcation leads to oscillations in the peak of the bump solution. We show a particular example of this oscillatory behavior in Figure \ref{fig:1d_dynamics}B.


\subsubsection{Normal Form for the Hopf Bifurcation on the Ring}
We wish to analyze the bifurcation to a \ypre{sloshing pulse} for the general integral equation:
\begin{equation}
\label{eq:th}
\frac{d\theta}{d\tau} = -q J(\theta) - g \int_0^\infty e^{-s} H(\theta(\tau-s)-\theta(\tau))\ ds
\end{equation}
as $g$ increases.  For simplicity, we will assume $J(\theta)$ is an odd periodic function (as is the case for $H(\theta)$ and through suitable rescaling of $g,q$, we will assume:
\begin{eqnarray*}
J(\theta) &=& \theta + j_3 \theta^3 + \ldots \\
H(\theta) &=& \theta + h_3 \theta^3 + \ldots.
\end{eqnarray*}
We also assume $q>0$ so that $\theta=0$ is stable without adaptation.  If, we use $H(\theta)=J(\theta)=\sin(\theta)$, then $j_3=h_3=-(1/6).$ The linearization about $\theta=0$ has the form:
\[
\theta_\tau = -q \theta - g \int_0^\infty e^{-s} (\theta(\tau-s)-\theta(\tau))\ ds
\]
which has the general solution, $e^{\lambda \tau}$. After some simplification, we find that 
\[
\lambda^2 +(1+q-g)\lambda+ q=0
\]
so there is an imaginary eigenvalue, $i\sqrt{q}:=i\omega$ when $g=1+q\equiv g_0$, so we expect a Hopf bifurcation will occur.

\begin{figure}[h!]
 \includegraphics[width=\textwidth]{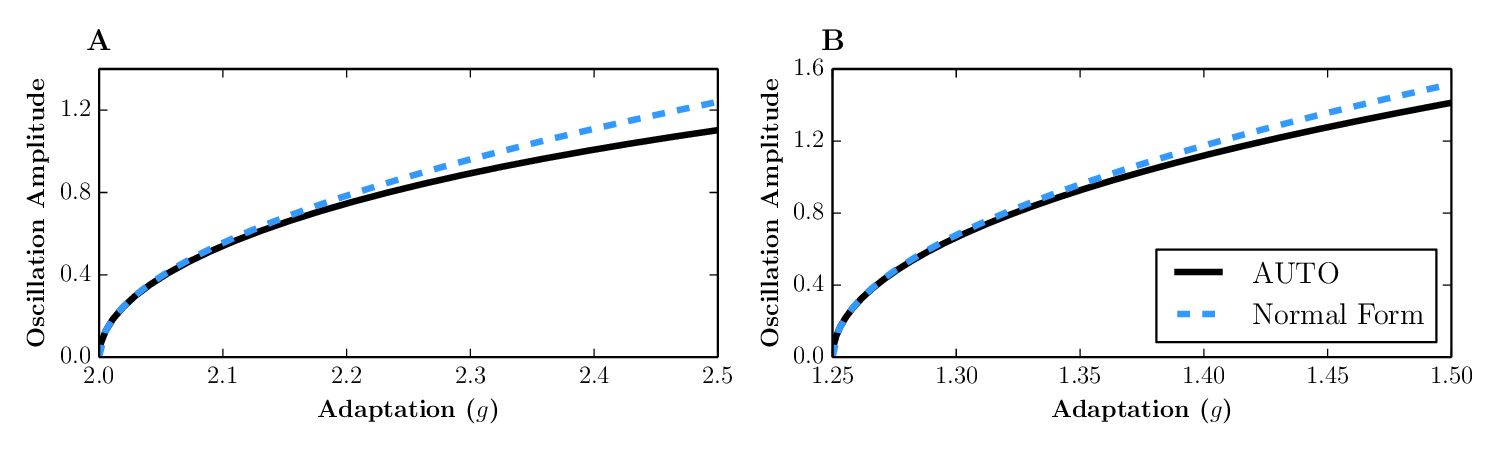}
 \caption{Normal form calculation for the neural field model on the ring. A: Amplitude of oscillations predicted by the normal form calculation (dashed blue) compared to the actual amplitude (solid black). $q=1$. B: Amplitude of oscillations predicted by the normal form calculation (dashed blue) compared to the actual amplitude (solid black). $q=0.25$.}\label{fig:1d_normal_form}
\end{figure}

\ympre{All nonlinearities are odd, so we can assume the multiple timescale expansion
\[
g= g_0 + \delta^2 g_2, \qquad \theta = \delta \theta_1(\zeta,\xi) + \delta^3 \theta_3(\zeta,\xi),
\]
where $\delta$ is the amplitude of the bifurcating solution, $\zeta= \tau$ is a ``fast'' time, and $\xi = \delta^2 \tau$ is a ``slow'' time. We detail the remaining steps of the normal form analysis in Appendix \ref{a:normal} and jump to the conclusion,}
\begin{equation}
\label{eq:normform}
\alpha \frac{d z}{d\xi} = z [\hat{\gamma}_0 + \hat{\gamma}_3 |z|^2]
\end{equation}
where
\begin{align*}
\alpha &= 1 - \frac{g_0}{1+2i\omega-\omega^2} = \frac{2}{1+q}( q + \sqrt{q}i) \\
\hat{\gamma}_0 &= g_2 \frac{i\omega}{1+i\omega} = \frac{g_2}{1+q}(q + \sqrt{q}i) \\
\hat{\gamma}_3 &= \frac{3q}{4q+1}\left[[q(12h_3-4j_3)-j_3]+i 18h_3 \sqrt{q}\right].
\end{align*} 
To get the actual normal form, we divide (\ref{eq:normform}) by $\alpha$, to obtain:
\[
\frac{dz}{d\xi} = z (g_2/2 + \gamma_3 |z|^2)
\]
where 
\[
\gamma_3 = \frac{3}{8q+2}\left[q (12qh_3-4qj_3-j_3+6h_3) + i \sqrt{q} (6qh_3-4qj_3-j_3)\right].
\]
If we assume that $j_3=h_3$ as would be the case if the input was the bump, itself, then
\[
\gamma_3 = h_3\frac{3 q (8 q+5)}{8q+2} - i h_3 \frac{3 \sqrt{q}(2q-1)}{8q + 2}.
\]
We compare the normal form calculation to the numerics in Figure \ref{fig:1d_normal_form}. We use \texttt{XPPAUTO} to compute the numerical bifurcation diagram. As expected, the normal form approximation is quite accurate near the bifurcation.

\subsection{Non-Constant Velocity Bump Solution on the Ring}
When adaptation is made even stronger, the solution breaks free from the oscillating state and travels across the periodic domain (Figure \ref{fig:1d_dynamics}C. The onset is shown numerically in the bifurcation diagrams of Figures \ref{fig:full_bifurcations} and \ref{fig:phase_bifurcations}, purple LP2). Due to the pinning term, the velocity of the bump is nonconstant.

\be{\bf{Remark.} As $q\to0$, we see in figure \ref{fig:phase_bifurcations}B that all the two-parameter curves converge to the point $g^*$ which is the point of onset of the traveling bumps with no input stimulus.}

\subsection{Chaos on the Ring}
\begin{figure}[h!]
 \includegraphics[width=\textwidth]{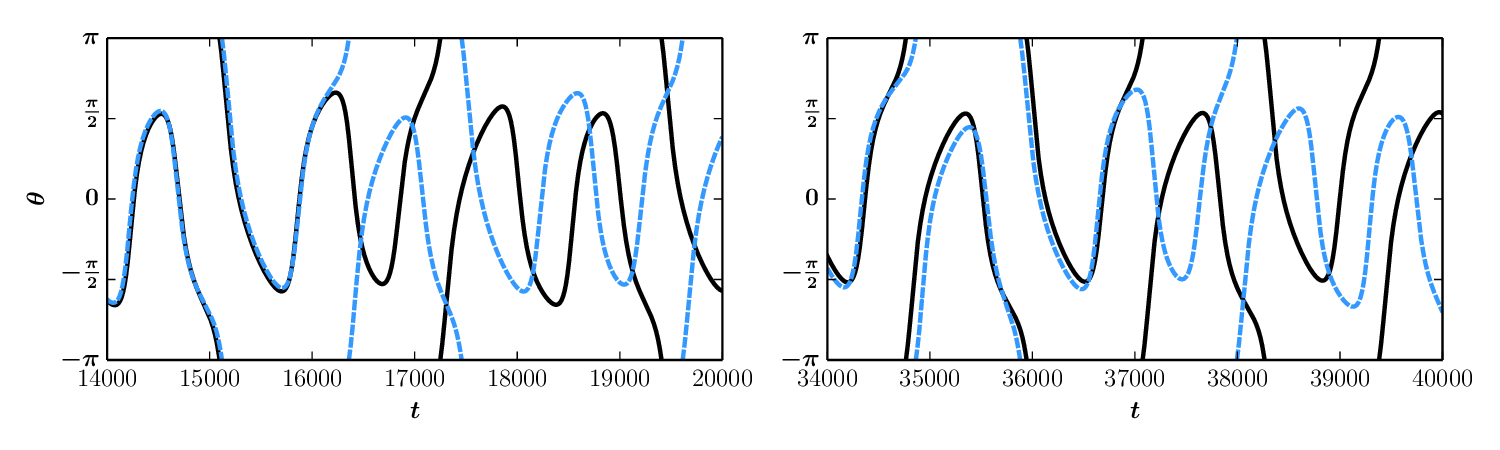}
 \caption{Chaotic dynamics of the traveling bump in the full neural field model (left) and the reduced phase model (right) on the ring. Original solutions are shown in black. Solutions with a different initial condition is shown in dashed blue. For each panel, initial conditions differ by 1e-7. A: $g=2.65,q=0.5$. B: $g=2.661,q=0.5$. For all simulations in this figure, $\ve = 0.01$.}\label{fig:1d_chaos}
\end{figure}

With $g,q>0$, there exists a small parameter range in which the neural field exhibits chaotic movement about the \ypre{ring}. Examples of this behavior are shown in Figure \ref{fig:1d_chaos}. In both panels, the initial conditions differ by 1e-7. The solutions in each panel remain nearly identical for a long time (we have truncated a significant portion of the simulation).

This section completes our analysis of the one-dimensional case. \be{ We have found a good match between the phase-reduced equations and the full neural model. For a fixed amplitude of the external input, we find a transition from a stationary bump to ``sloshers'', and, finally to bumps that move completely around the ring, modulated traveling bumps.}   In the sections to follow, we repeat the analytical and numerical analysis for the two-dimensional domain.


\section{Torus Domain}
\ypre{In this section, we define the domain $\Omega$ as the torus, or the square $[-\pi,\pi]\times[-\pi,\pi]$ with periodic boundary conditions. We seek to analyze the full neural field model on this domain using the same bifurcation analysis performed in the one-dimensional case. We begin by considering simplifications that allow us to use standard bifurcation analysis tools like \texttt{XPPAUTO}.}


\subsection{Approximation of the Neural Field Model on the Torus}
\bge{In order to numerically investigate the full neural field equation (\ref{eq:u1}-\ref{eq:z1}), we need to either discretize space in two-dimensions or use an approximation of the kernel that is degenerate. (Since the integral operator is compact, it can be approximated to arbitrary precision by a degenerate integral operator ;see section 2.8 \cite{pipkin}). Thus}
to study the dynamics of the full neural field model on a two-dimensional domain, we take a Fourier truncation of the kernel to make the integral in Equation \eqref{eq:ufull} separable.  This truncation allows us to rewrite the infinite dimensional system as a finite system of ODEs and use traditional dynamical systems tools like \texttt{XPPAUTO} to analyze the system. To begin, take the truncated Fourier approximation to the kernel,
\begin{equation}\label{eq:kernel_trunc}
 K(x,y) = k_{00} + k_{10}\cos(x) + k_{01}\cos(y) + k_{11}\cos(x)\cos(y),
\end{equation}
and plug it into Equation \eqref{eq:ufull}:
\begin{equation*}
 u(\x) = \io K(x_1-y_1,x_2-y_2) f(u(\y))d\y.
\end{equation*}
After expanding the kernel using standard trigonometric identities, we derive the time-varying solutions
\begin{equation}\label{eq:u_trunc}
 \begin{split}
 u(\x,t) &= a_{00}(t) + a_{10}(t)\cos(x_1)  + a_{01}(t)\cos(x_2) \\
 &\quad + b_{10}(t)\sin(x_1) + b_{01}(t)\sin(x_2) \\
 &\quad + a_{11}(t)\cos(x_1)\cos(x_2) + b_{11}(t)\sin(x_1)\sin(x_2)\\
 &\quad + c_{1}(t)\sin(x_1)\cos(x_2) + c_{2}(t)\cos(x_1)\sin(x_2).\\
 z(\x,t) &= E_{00}(t) + E_{10}(t)\cos(x_1) + E_{01}(t)\cos(x_2)\\
 &\quad + F_{10}(t)\sin(x_1)  + F_{01}(t)\sin(x_2) \\
 &\quad + E_{11}(t)\cos(x_1)\cos(x_2) + F_{11}(t)\sin(x_1)\sin(x_2)\\
 &\quad + G_{1}(t)\sin(x_1)\cos(x_2) + G_{2}(t)\cos(x_1)\sin(x_2).
 \end{split}
\end{equation}
where the coefficients satisfy
\begin{equation}\label{eq:u_trunc_coeffs}
\begin{split}
 a_{ij}' &= -a_{ij} + k_{ij}p_{ij}(t) + \ve (qu_{ij} - gE_{ij}),\\
 b_{ij}' &= -b_{ij} + k_{ij}r_{ij}(t) - \ve gF_{ij},\\
 c_{i}' &= -c_{i} + k_{11}s_{i}(t) - \ve g G_{i},\\
 \xi' &= \ve\beta(-\xi+\zeta),
\end{split}
\end{equation}
where $i=0,1$ and $j=0,1$. The dummy variables $\xi,\zeta$ represent each of the pairs $(a_{ij},E_{ij}), (b_{ij},F_{ij})$, and $(c_i,G_i)$.  The time-varying functions $p_{ij},r_{ij},s_i$ are defined as
\begin{align*}
\begin{array}{ll}
 p_{00}(t) = \io f(u(\y,t))d\y & p_{01} = \io \cos(y_2)f(u(\y,t))d\y\\
 p_{10}(t) = \io \cos(y_1) f(u(\y,t))d\y & p_{11} = \io \cos(y_1)\cos(y_2)f(u(\y,t))d\y\\
  & \\
 r_{01}(t) = \io \sin(y_2)f(u(\y,t))d\y & r_{10} = \io \sin(y_1)f(u(\y,t))d\y\\
 r_{11}(t) = \io \sin(y_1)\sin(y_2) f(u(\y,t))d\y\\
  & \\
 s_{1}(t) = \io \cos(y_1)\sin(y_2) f(u(\y,t))d\y & s_{2} = \io \sin(y_1)\cos(y_2) f(u(\y,t))d\y,
\end{array}
\end{align*}
and the coefficients $u_{ij}$ are taken from the truncated Fourier series of the steady-state solution,
\begin{equation*}
 u_0(x,y) = u_{00} + u_{10}\cos(x) + u_{01}\cos(y) + u_{11}\cos(x)\cos(y).
\end{equation*}
The coefficient values of the kernel and steady-state bump solutions are shown in Tables \ref{tab:k_fourier} and \ref{tab:u_fourier}, respectively. Details on how we approximate the spatial integrals of $p_{ij},r_{ij},s_i$ are in Appendix \ref{a:numerical_int}.

\begin{figure}[h!]
\centering
\includegraphics[width=.75\textwidth]{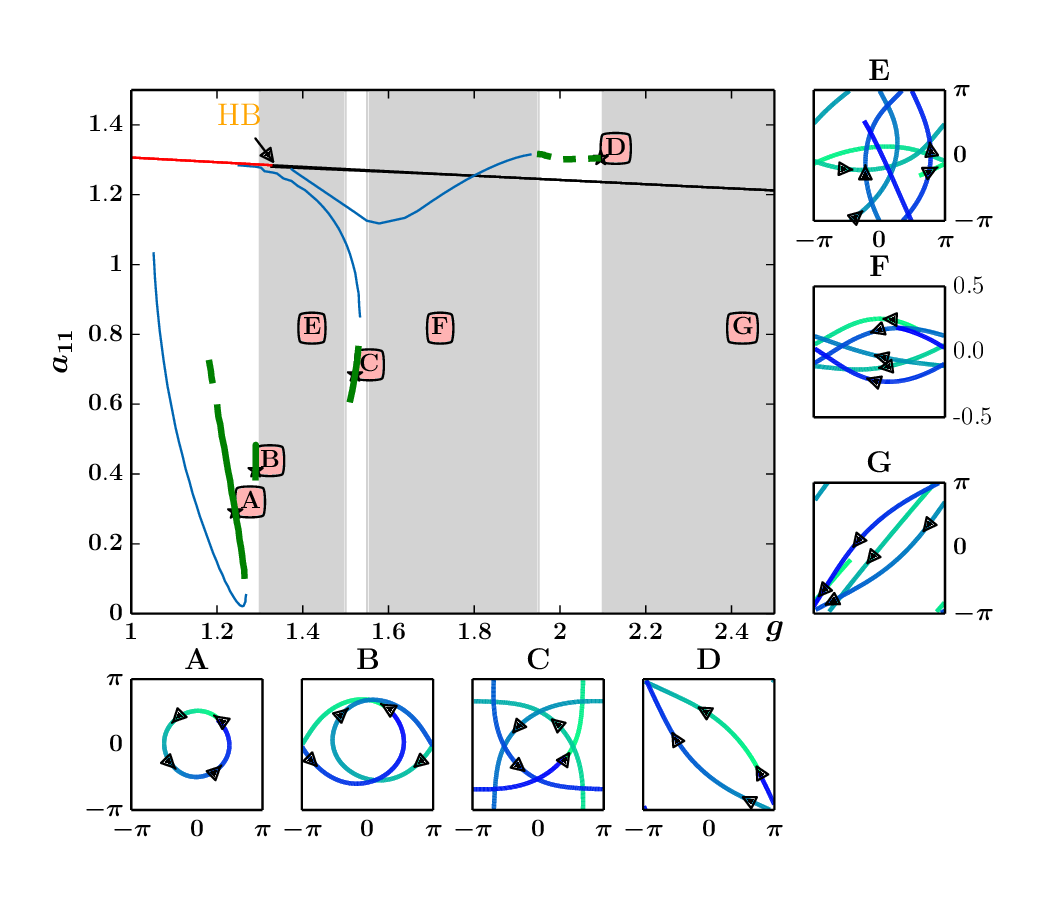}
 \caption{Bifurcation diagram of the truncated neural field model \ypre{on the torus with $g$ as a bifurcation parameter and $q=0.1$}. The stable fixed point (red line) undergoes a subcritical Hopf bifurcation (HB, orange) and becomes an unstable fixed point (black line). The green and blue lines represent stable and unstable oscillations, respectively. Thick, solid green lines represent stable oscillations. Thick, dashed green lines represent stable oscillations that wrap around the torus. Thin solid blue lines represent unstable periodic solutions. \ypre{Stable attractors are shown in panels A--D. In panels E--G, we show solutions in parameter regimes without stable periodic attractors. These solutions are displayed in a relatively short time window after integrating for long times and travel from light to dark. In panel E ($g=1.4$), we integrate for $t=5000$ time units and show the last $30\%$ of the data. In panel F ($g=1.7$), we integrate for $t=8000$ time units and show the last $10\%$ of the data. In panel G ($g=2.4$), we integrate for $t=8000$ time units and show the last $6\%$ of the data. We initialize the solutions of panels E--F using standard normally distributed random variables.} Parameter $\ve = 0.01$.}\label{fig:full_q=.1}
\end{figure}

We show the bifurcation diagram and salient solutions of this system in Figure \ref{fig:full_q=.1}. The only bifurcations classified by AUTO are a subcritical Hopf bifurcation (HB, orange) and limit points (limit points occur at each change in stability of periodic solutions). The Hopf bifurcation leads to small amplitude unstable solutions. We attribute the existence of a subcritical Hopf bifurcation to the coarse discretization of the spatial domain in the parameters of Equation \eqref{eq:u_trunc_coeffs}. Although not shown here, a finer discretization of the spatial domain using 200 intervals results in a qualitatively supercritical Hopf bifurcation.

To summarize, we find the usual types of oscillatory solutions in this truncated neural field model as we found in the neural field model. A stable limit cycle of this system is shown in the first bottom left panel (A, which corresponds to the initial conditions taken from the point A in the bifurcation diagram). A large-sloshing solution exists for slightly larger $g$ (B,C), and eventually, for sufficiently large $g$, there exist only traveling bump solutions D. \ypre{Additional non-periodic attractors are shown in panels E--G. The attractors shown in this figure are simply those with the largest basins of attraction. Generally, starting random initial conditions with $g$ anywhere in the range $1.55 < g < 1.95$ (the gray shaded area labeled F in the main plot) results in solutions that qualitatively match panel F. The same holds for the shaded areas E and F, with their corresponding panels. Indeed, there exist several attractors not shown in this figure that are more difficult to find numerically. However, the focus of this study is not the thorough classification of attractors in the truncated neural field model, so we move on to the analysis of the phase model on the torus.}

\begin{figure}[h!]
\centering
\includegraphics[width=.75\textwidth]{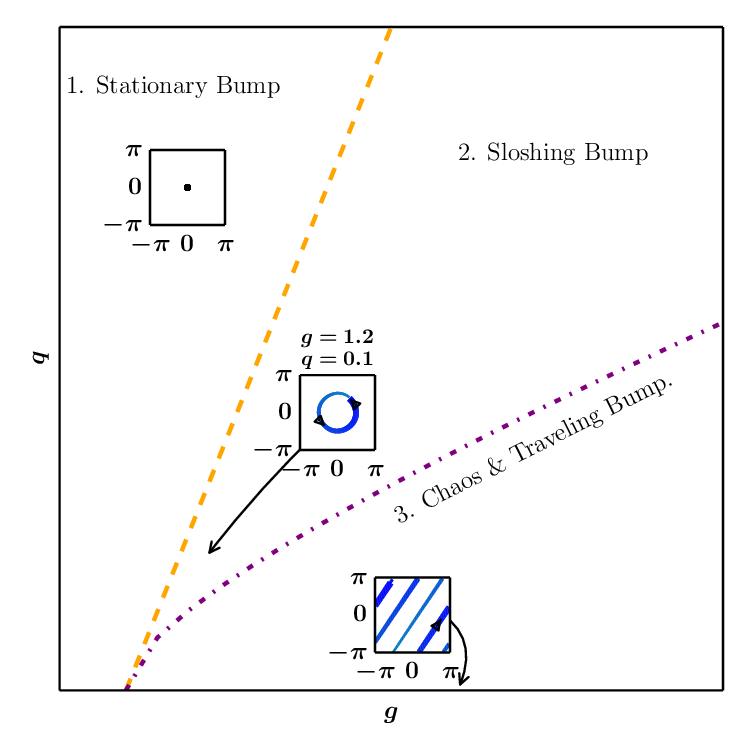}
 \caption{\ypre{Cartoon of the parameter space of the approximate neural field model on the torus (Equation \eqref{eq:u_trunc_coeffs}). The most salient solutions are shown. Solutions advance in time from light to dark, thin to thin. For sufficiently small $g$ or sufficiently large $q$, the bump solution tends to a stationary solution. By increasing $g$ or decreasing $q$ to $g=1.2,q=0.1$, the centroid of the bump solution oscillates about the origin. For larger $g$, say $g=4,q=0.5$, the solution begins to traverse chaotically about the domain. When $q=0$, there exists a constant velocity traveling bump solution for $g$ sufficiently large, e.g., $g=3$.} Parameter $\ve = 0.01$.}\label{fig:2dfull2par}
\end{figure}

\subsection{Approximations of the Phase Model on the Torus}
We now turn to the analysis of the phase dynamics in two-dimensions and begin by reducing the number of parameters with the same rescaling used to obtain Equation \eqref{eq:phs_rscl} in the one dimensional case,
\begin{equation}\label{eq:phs_rscl2}
\frac{d\theta_i}{d\tau} = qJ_i(\q) -g\int_0^\infty e^{-s} H_i(\q(\tau-s)-\q(\tau))ds, \quad i=1,2,
\end{equation}
and recall that
\begin{equation*}
 H_i(\q) = \io f'(u_0(\x))\pa_i u_0(\x) u_0(\x+\q)\ d\x.
\end{equation*}
Details on the numerical integration of the integro-differential equation, Equation \eqref{eq:phs_rscl2}, are shown in Appendix \ref{a:numerical_int}.

To facilitate the study of existence and stability of solutions, we consider two approximations to $H_i$ to be used in Equation \eqref{eq:phs_rscl2}: the first is a high-accuracy Fourier series of $H_i$, and the second is a low-accuracy Fourier series of $H_i$. We detail these approximations in turn.

For the accurate Fourier series approximation of $H_i$, we use one of two equivalent forms
\begin{equation}\label{eq:fourier_full1}
 H_1(\q) = \sum_{n,m\in\mathbb{Z}} a_{nm} \sin(n\theta_1)\cos(m\theta_2),
\end{equation}
where, due to the odd (even) property of the first (second) coordinate, the coefficients have the property that $a_{n,\pm m} = - a_{-n,\pm m}$. We can then rewrite this Fourier series into the equivalent form,
\begin{equation}\label{eq:fourier_full2}
 H_1(\q) = \sum_{n,m\in\mathbb{Z}} \hat a_{nm} \sin(n\theta_1 + m\theta_2),
\end{equation}
where $\hat a_{nm} = 4a_{nm}$. This equivalent form makes integrals much easier to compute. We use both forms interchangeably as we see fit, and abuse notation in Equation \eqref{eq:fourier_full2} by removing the hats from the coefficients. We find that 30 Fourier coefficients provides a sufficiently good approximation for simulations on a $64\times 64$ domain (the error is on the order of 1e-7).

For the low-accuracy Fourier series, we consider a more substantial truncation of the interaction function using only 3 Fourier coefficients. While this truncation is drastic, it allows us to analyze Equation \eqref{eq:phs_rscl2} more rigorously. We derive the 3 term $H_i$-function starting with the same Fourier truncation of the kernel as above, which leads to the same steady-state bump solution,
\begin{equation*}
 u_0(x,y) = u_{00} + u_{10}\cos(x) + u_{01}\cos(y) + u_{11}\cos(x)\cos(y),
\end{equation*}
which in turn leads to a truncated $H_i$ function,
\begin{equation*}
 H^F_1(\theta_1,\theta_2) =  \sin(\theta_1)(h_{10} + h_{11}\cos(\theta_2)),
\end{equation*}
where
\begin{align*}
h_{10} &= 4u_{10}\io \sin^2(x)(u_{10} + 2\cos(y) u_{11}) dxdy,\\
h_{11} &= 8u_{11}\io \cos(y)\sin^2(x)(u_{10} + 2\cos(y) u_{11})dxdy.
\end{align*}
For simplicity, analysis of this $H_i$ function uses the simpler form
\begin{equation}\label{eq:h_trunc}
 H^F_1(\theta_1,\theta_2) = \sin(\theta_1)(1 + b\cos(\theta_2)),
\end{equation}
where we have absorbed $h_{10}$ into the parameter $g$ of Equation \eqref{eq:phs_rscl2}, and relabeled $h_{11}/h_{10}$ as $b$. Naturally, it follows that $H_2^F(x,y) = H_1^F(y,x)$ and $H_i^F = -J_i^F$. 

Using the truncated interaction function $H_1^F$ enables us to use traditional dynamical systems tools and techniques to identify qualitative dynamics of Equation \eqref{eq:phs_rscl2} through a bifurcation analysis. 

Finally, for $H_i$ and all of its approximations, we require the following properties to hold:
\begin{enumerate}
 \item $\pa H_1(0,0)/\pa y = \pa H_2(0,0)/\pa x = 0$,
 \item $\pa H_1(0,0)/\pa x,  \pa H_2(0,0)/\pa y > 0$,
 \item $\pa J_1(0,0)/\pa x,  \pa J_2(0,0)/\pa y < 0$.
\end{enumerate}
\be{Properties 1 and 2 follow from the evenness of the $\ve=0$ bump solution and 3 is made WLOG since we could just change the sign of $q$ otherwise.} 

To summarize, we consider two approximations to $H_i$:
\begin{align*}
 H_1(\q) &= \sum_{n,m=1}^{30} a_{nm}\sin(n\theta_1)\cos(m\theta_2) \propto\sum_{n,m=1}^{30} a_{nm} \sin(n\theta_1 + m\theta_2),\\
 H_1^F(\q) &= \sin(\theta_1)(1 + b\cos(\theta_2)).
\end{align*}
\ypre{We note that the second approximation, $H_1^F$, is a result of using the truncated kernel in Equation \eqref{eq:kernel_trunc}}.

\subsubsection{Equivalent Truncated Phase Model on the Torus}
For the truncated function $H_i^F$, we transform the delay integro-differential equations into a system of ordinary differential equations using identical arguments used to transform the phase equation on the ring from a delay integro-differential equation into a system of ODEs. The new system is
\begin{align}\label{eq:phase_trunc}
\begin{split}
\theta_i' &= qJ_i^F(\q)-g(\eta_{i1}+\eta_{i2}),\quad i=1,2\\
N'&=-N+P\\
M'&=-M+Q
\end{split}
\end{align}
where 
\begin{align*}
(N,P) &\in \{(cx,\cos\theta_1),(cy,\cos\theta_2),(sx,\sin\theta_1),(sy,\sin\theta_2)\}\\
(M,Q) &\in \{(sxsy,\sin(\theta_1)\sin(\theta_2)),(sxcy,\sin(\theta_1)\cos(\theta_2)),\\
  &\quad\quad(cxsy,\cos(\theta_1)\sin(\theta_2)),(cxcy,\cos(\theta_1)\cos(\theta_2)) \}\\
\eta_{11}&=sx\cos(\theta_1)-cx\sin(\theta_1)\\
\eta_{12}&=b[\cos(\theta_1)\cos(\theta_2)sxcy - \sin(\theta_1)\cos(\theta_2)cxcy\\
   &\quad\quad+ \cos(\theta_1)\sin(\theta_2)sxsy - \sin(\theta_1)\sin(\theta_2)cxsy].
\end{align*}
The function $\eta_{21}$ ($\eta_{22}$) is the same as $\eta_{12}$ ($\eta_{11}$) with $\theta_1$ and $\theta_2$ flipped and each $x$ and $y$ flipped in the notation (for example, $sxcy$ and $\cos(\theta_2)$ in $\eta_{12}$ become $sycx$ and $\cos(\theta_1)$ in $\eta_{21}$, respectively).

\begin{figure}[h!]
\centering
\includegraphics[width=.7\textwidth]{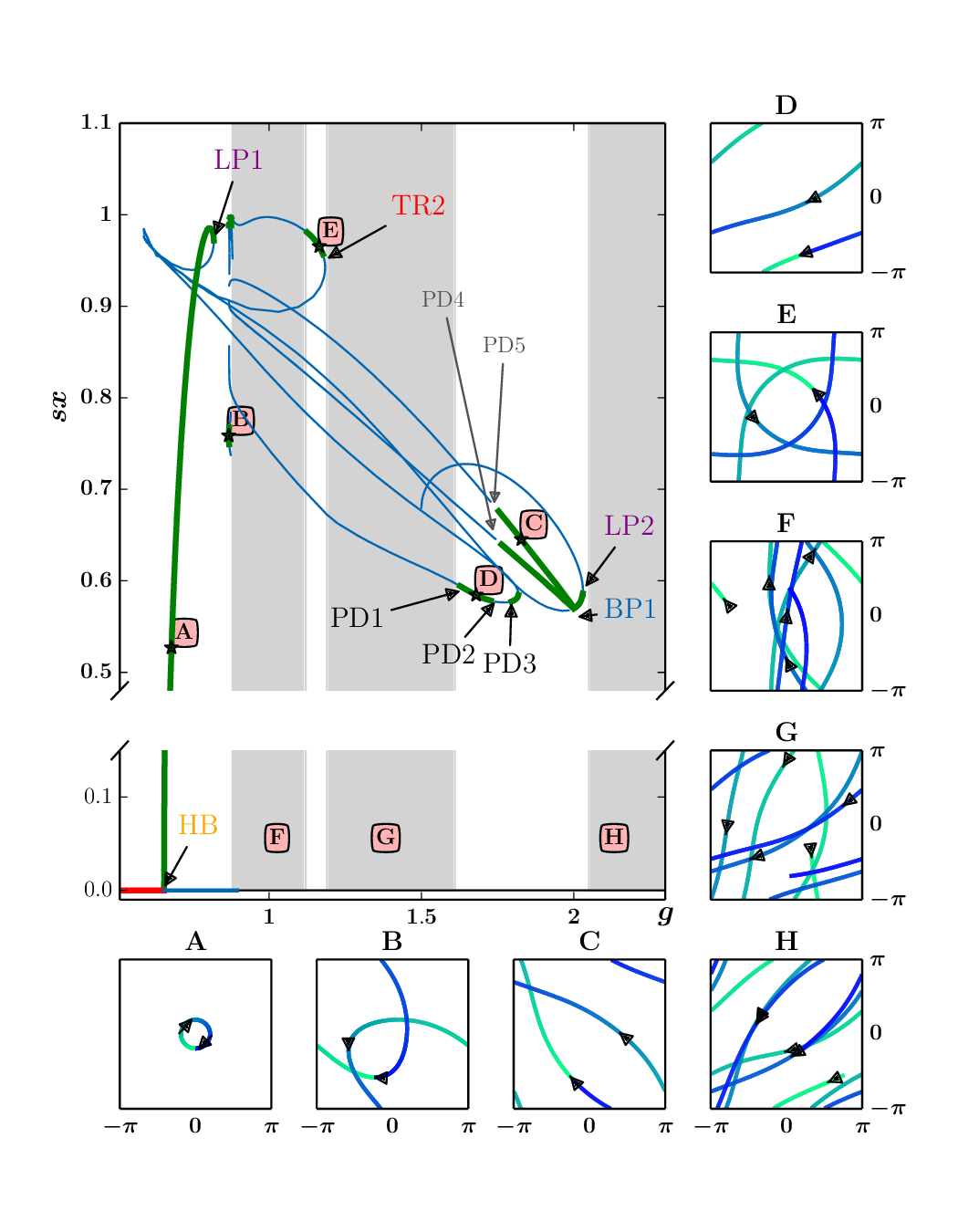}
 \caption{Bifurcation diagram of the equivalent truncated phase model on the torus over varying values of $g$ with $q=0.1$. Some branches refined using \texttt{XPPY}\cite{xppy}. Sample solutions (labeled A--H in the bifurcation diagram) are shown in the subplots to the bottom and right of the diagram. Bifurcations are labeled according to the type: Hopf (HB), limit point (LP), torus (TR), period-doubling (PD), and branch point (BP). The number following each bifurcation type correspond to the same bifurcation type and number in the two parameter bifurcation in Figure \ref{fig:2d2par}. \ypre{Panels A--E show stable attractors.  In panels F--H, we show solutions in parameter regimes without stable periodic attractors. These solutions are displayed in a relatively short time window after integrating for long times and travel from light to dark. In panel F ($g=1.05$), we integrate for $t=500$ time units and show the last $9\%$ of the data. In panel F ($g=1.5$), we integrate for $t=500$ time units and show the last $7\%$ of the data. In panel G ($g=2.15$), we integrate for $t=500$ time units and show the last $5\%$ of the data. We initialize the solutions of panels F--H using standard normally distributed random variables.} Parameter $\ve = 0.01$.}\label{fig:q=.1}
\end{figure}

We show the many bifurcations and salient solutions of this system in Figure \ref{fig:q=.1}. We find that there exists a stable sloshing bump solution that arises from a Hopf bifurcation (solution A, bifurcation HB). Due to the symmetry of the system, there is also an unstable sloshing solution in an axial direction (G) that arises from the same Hopf bifurcation. For slightly larger parameter values, there is bistability of large-sloshing solutions (H and B), and an even larger-sloshing solution (F) that loses stability through a torus bifurcation (TR). For this choice of $q=0.1$, the solutions are chaotic for parameter values between the first torus bifurcation (TR1) and the first period doubling bifurcation (PD1). The multitude of period doubling bifurcations beyond this point represents the onset of chaotic behavior of the system due to the Fourier truncation, the error of which is proportional to $g$. This error is more apparent beyond limit point LP2 where there are no more traveling bump solutions, which qualitatively disagrees with the original system where traveling bump solutions exist for even relatively large $g$.

\begin{figure}[h!]
\centering
\includegraphics[width=.75\textwidth]{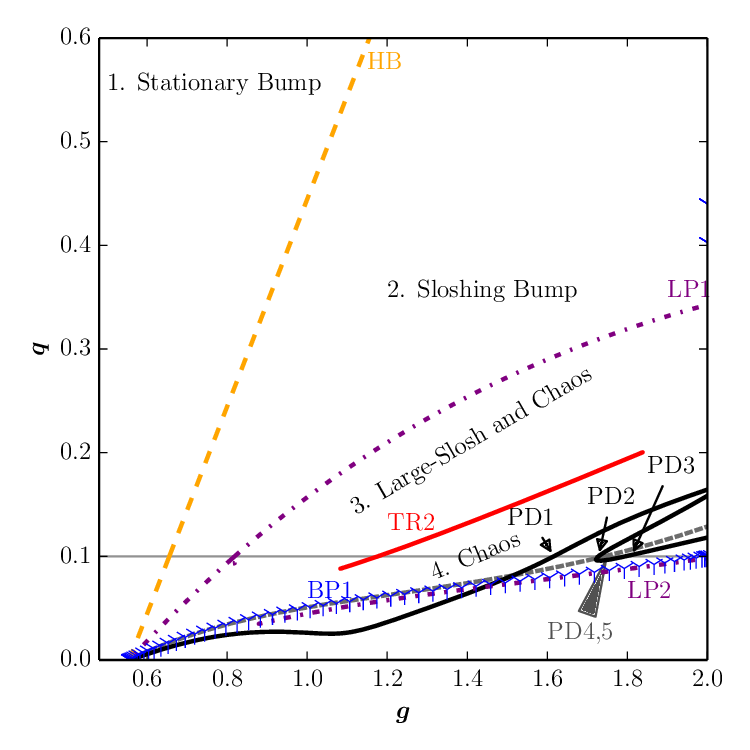}
 \caption{Two parameter bifurcation diagram of the equivalent truncated phase model on the torus. The parameter regions are separated into stationary solutions (1.), sloshing solutions (2.), large-sloshing solutions (3.), and generally chaotic solutions (4.). To the right of the curve LP2 (purple dashed) for $g\geq 1.5$, the qualitative behavior breaks down as this bifurcation point marks the end of traveling bump solutions. Parameter $b=0.8$.}\label{fig:2d2par}
\end{figure}

\ypre{The most salient} bifurcations are captured in the two-parameter bifurcation in Figure \ref{fig:2d2par}. There are several qualitative similarities to the two parameter bifurcation diagram of the phase model on the ring. In particular the transition from the stationary bump to the sloshing bump, and from sloshing to large-sloshing. However, for the two-dimensional domain, there are much larger regions of chaotic solutions. 

\ypre{In the following sections, we study the dynamics of the original phase model and the truncated phase model and repeat most of the analysis as completed in the ring domain. In particular, using a combination \ympree{of} numerical and analytical methods, we analyze the existence and stability of traveling bump solutions, and the existence of a Hopf bifurcation.}

\subsection{Constant Velocity Bump Solution on the Torus}
\begin{figure}[h!]
    \centering
    \includegraphics[width=\textwidth]{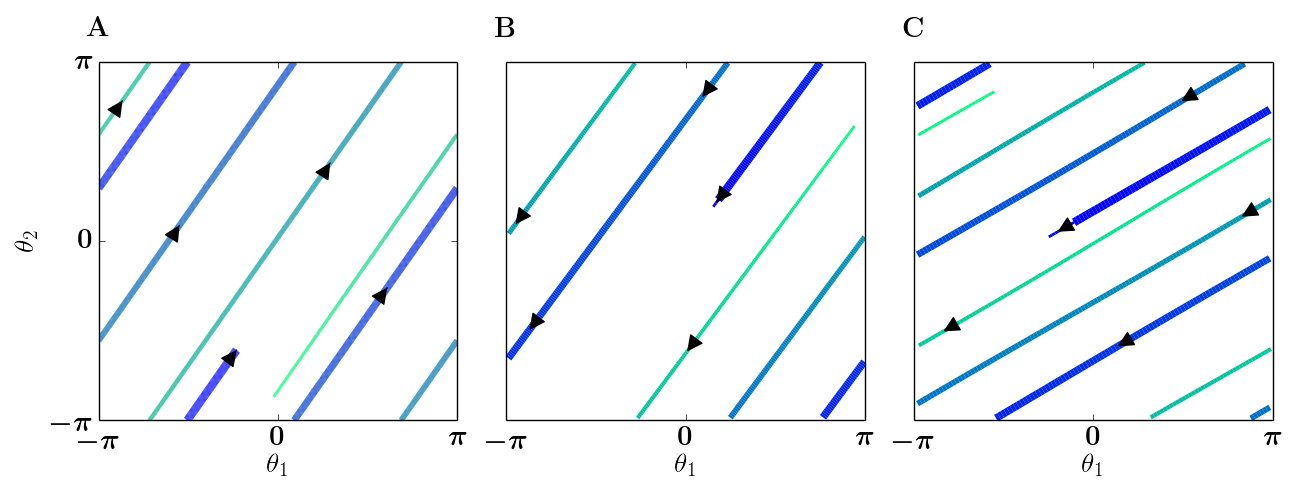}
    \caption{Constant velocity dynamics of the traveling bump on the torus. \ypre{The curve that goes from light to dark and thin to thick represents the movement of the centroid over time.} A: Full neural field model on the torus, $q=0,g=3$, \ypre{simulated for $t=7000$ time units with the last 60\% of the data shown}. B: Phase model on the torus with the accurate Fourier series of $H_i$, $q=0,g=2.2$, \ypre{simulated for $t=6,700$ time units with the last 10\% of the data shown}. C: \ypre{Phase model on the torus with the truncated Fourier series $H_i^F$, $q=0,g=2.5$, simulated for $t=3,500$ time units with the last 20\% of the data shown}. For these parameter choices, the axial directions are unstable and over long times converge to non-axial directions. Parameter $\ve = 0.01$.}\label{fig:2d_const}
\end{figure}

In this section, we analyze the existence and stability of constant velocity bump solutions on the torus \be{for $q=0$, the only case in which there can be constant velocity traveling bumps.} Figure \ref{fig:2d_const} shows the type of solutions we analyze in this section: constant velocity traveling bump solutions in the full neural field model (panel A), the reduced model with the accurate Fourier approximation (panel B), and the truncated reduced model (panel C).

\subsubsection{Existence}
To show existence of solutions in the axial directions, we only need to show existence of the solution $\theta_1(\tau) = \nu\tau$ and $\theta_2(\tau) = 0$. We plug this ansatz into \eqref{eq:phs_rscl2} and rearrange to yield
\begin{equation}\label{eq:g_nu2}
 \ympree{g=\Gamma(\nu)} \equiv \frac{\nu}{\int_0^\infty e^{-s} H_1(\nu s, 0) ds}.
\end{equation}
The analysis of this equation is identical to the one-dimensional case, Equation \eqref{eq:g_nu}. \be{By varying $\nu$ from zero, we can find the values of $g$ where there are solutions.} \be{ Values of $g$  for which $\nu$ is nonzero imply there exists a traveling bump solution.} \ympree{In the case of the truncated $H$ function, we compute this integral explicitly to derive the velocity $\nu$ as a function of adaptation strength $g$:
\begin{equation*}
 \Gamma(\nu) = \frac{1+\nu^2}{1+b}.
\end{equation*}
To determine the critical value for the existence of axial constant velocity bump solutions, we take the limit $\lim_{\nu \rightarrow 0} \Gamma(\nu)$:
\begin{equation*}
g^* = \lim_{\nu \rightarrow 0} \Gamma(\nu) = \frac{1}{1+b}.
\end{equation*}
}

\begin{figure}[h!]
 \centering
 \includegraphics[width=.9\textwidth]{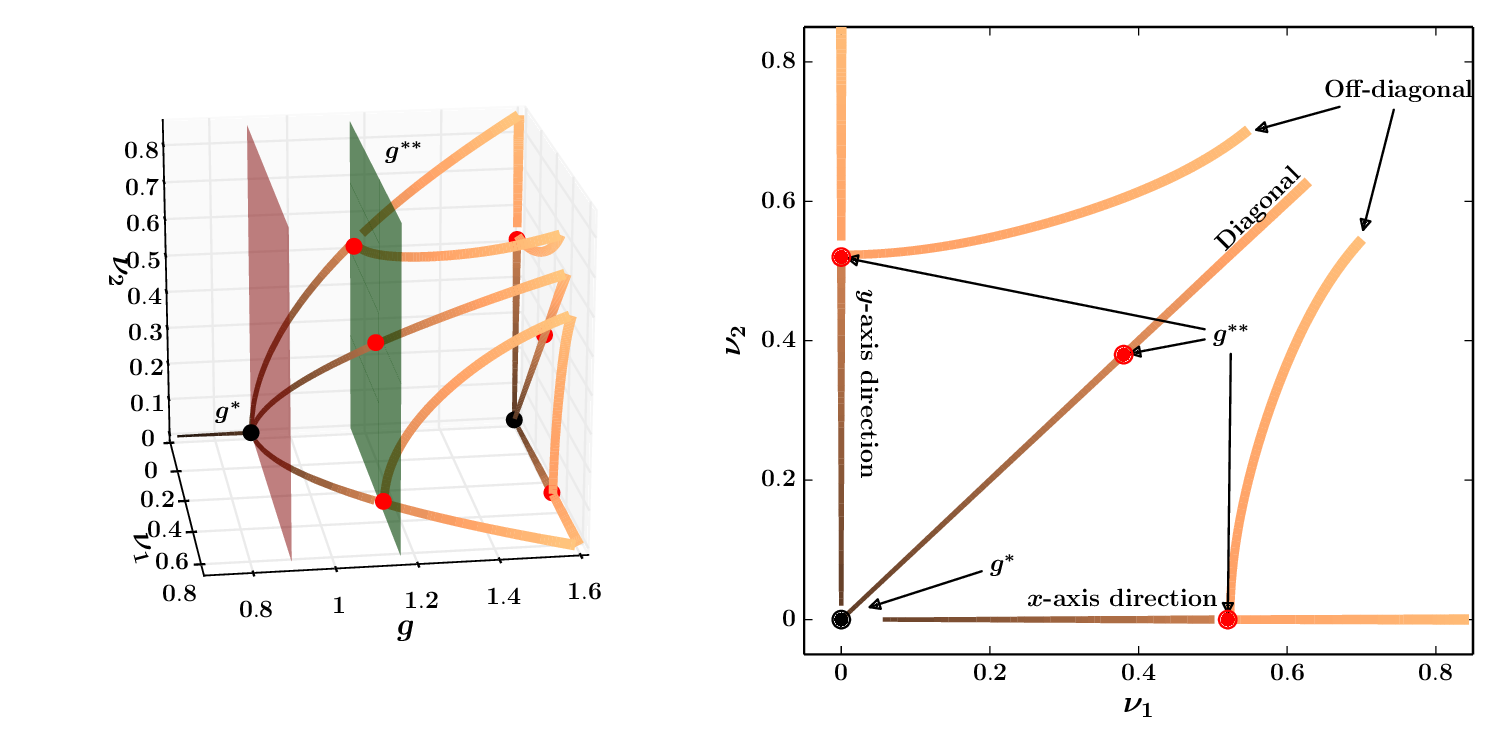}
 \caption{Existence of traveling bump solutions using the accurate approximation to the interaction function $H_i$. Left: After a first critical value \ympree{($g^*$)} of the bifurcation parameter $g$ (red plane), there exist traveling bumps in the axial directions. After a second critical value \ympree{($g^{**}$)} (marked by a green plane), off-diagonal solutions form and continue to persist for large $g$. The dark to light color gradient and thin to thick thickness gradient corresponds to increasing values of $g$. Right: The projection of the branches on the left onto the $g=1.6$ plane. A given point on one of these branches marks the magnitude and direction of a traveling bump. If necessary, one can approximate the parameter value $g$ of this traveling bump by looking at the thickness and color of the chosen point and looking back at the branches in the left panel.}\label{fig:twod_wave_exist}
\end{figure}

\begin{figure}[h!]
 \centering
 \includegraphics[width=.9\textwidth]{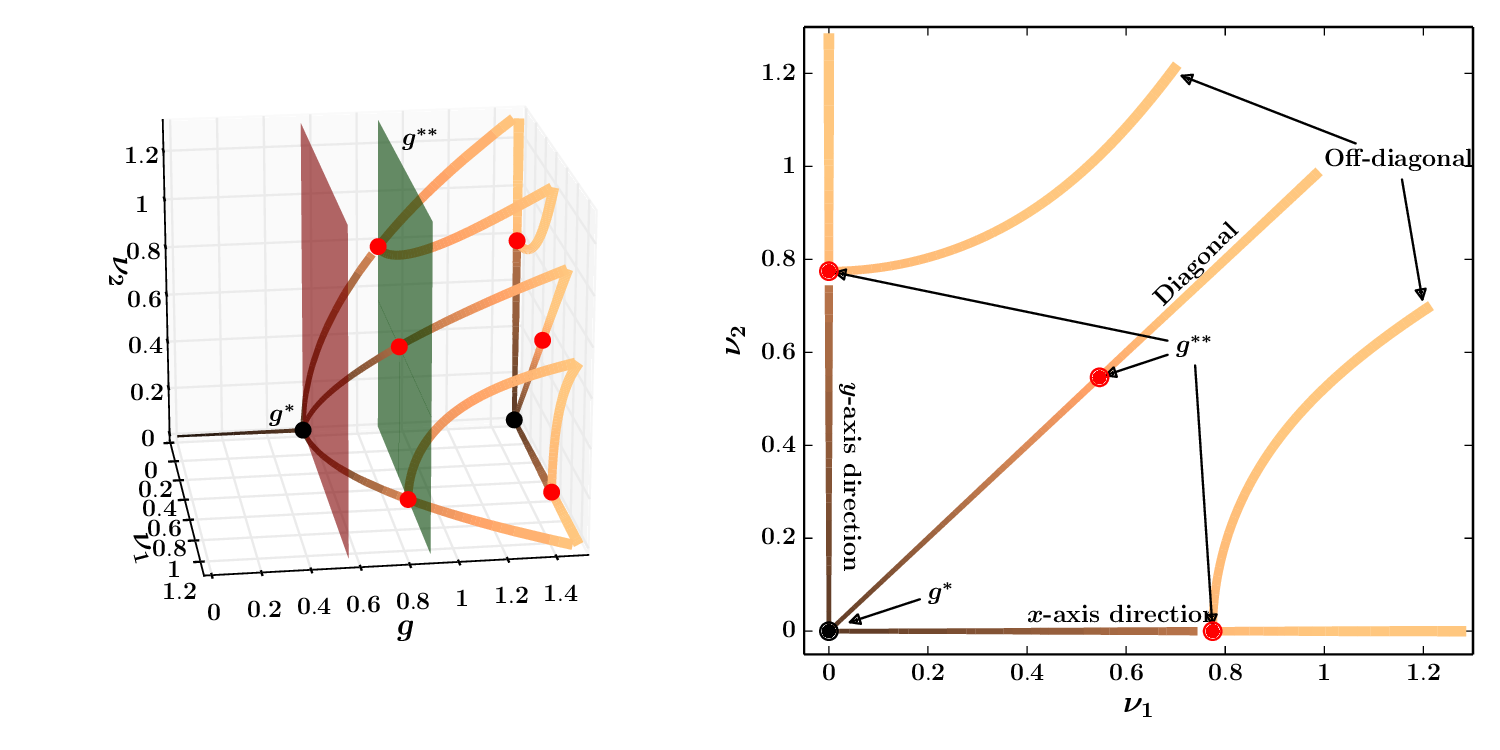}
 \caption{Existence of traveling bump solutions using the truncated interaction function $H_i^F$ ($q=0,b=0.8$). Left: After a first critical value \ympree{($g^*$)} of the bifurcation parameter $g$ (red plane), there exist traveling bumps in the axial directions. After a second critical value \ympree{($g^{**}$)} (marked by a green plane), off-diagonal solutions form and continue to persist for large $g$. The dark to light color gradient and thin to thick thickness gradient corresponds to increasing values of $g$. Right: The projection of the branches on the left onto the $g=4$ plane. A given point on one of these branches marks the magnitude and direction of a traveling bump. If necessary, one can approximate the parameter value $g$ of this traveling bump by looking at the thickness and color of the chosen point and looking back at the branches in the left panel.} \label{fig:twod_wave_exist_trunc}
\end{figure}

To show existence of non-axial solutions, we use the ansatz $\theta_1(\tau) = \nu_1\tau$ and $\theta_2(\tau) = \nu_2\tau$ where $\nu_1,\nu_2\neq 0$. There exist non-axial traveling bump solutions if $\nu_1,\nu_2$ simultaneously satisfy
\begin{equation}\label{eq:twod_wave_exist}
\begin{split}
 0 &= -\nu_1 + g G(\nu_1,\nu_2),\\
 0 &= -\nu_2 + g G(\nu_2,\nu_1),
\end{split}
\end{equation}
where
\begin{equation*}
 G(\nu_1,\nu_2) = \int_0^\infty e^{-s} H_1(\nu_1 s,\nu_2 s) ds.
\end{equation*}
\ympree{We can not compute the velocities $\nu_1,\nu_2$ explicitly as a function of $g$, but we can} exploit the Fourier series of $H_i$ to compute $G$ explicitly, allowing us to use \texttt{XPPAUTO} to follow the velocities as a function of the adaptation parameter $g$. The existence of traveling solutions using the accurate Fourier series is shown in Figure \ref{fig:twod_wave_exist}, and the existence of traveling solutions using the truncated Fourier series is shown in Figure \ref{fig:twod_wave_exist_trunc}.

\ympree{In these figures, we find that the truncated model (Figure \ref{fig:twod_wave_exist_trunc}) exhibits a similar set of traveling bump solutions as the full phase model (Figure \ref{fig:twod_wave_exist}). In particular, each system at a critical value $g^*$, bifurcates into two axial solutions and one diagonal solution. For larger $g$, the system bifurcates again at another critical value $g^{**}$, giving rise to two non-axial, non-diagonal constant velocity directions. Indeed, negative velocity solutions exist, but as these solutions are symmetric up to multiples of a 90-degree rotation about the $g$-axis, we only show the positive directions. \be{The ``mixed'' solutions that branch off for $g>g^{**}$ are, in general, not rationally related so that the resulting traveling bumps will densely cover the torus. As such quasi-periodic solutions are often not structurally stable, we expect to see complex and possibly chaotic behavior when $q>0$. Indeed, looking at Figure \ref{fig:2d2par}, we see that most of the complex behavior occurs for a small value of $q$ and $g$ sufficiently large.}
}

\ypre{Now that we have shown existence of traveling bump solutions, we proceed with a stability analysis.}

\subsubsection{Stability}
We begin this section with stability of traveling bump solutions in the axial directions. We perturb off the axial solution, $\theta_1(\tau) = \nu\tau + \ve e^{\lambda_1 \tau}$ and $\theta_2(\tau) = 0 + \ve e^{\lambda_2 \tau}$, with $\text{Re}(\lambda_i)>-1$. The first order terms yield two independent eigenvalue problems
\begin{align*}
 \lambda_1 &= -g \int_0^\infty e^{-s} \frac{\pa H_1}{\pa x}(\nu s, 0)[e^{-\lambda_1 s}-1]ds,\\
 \lambda_2 &= -g \int_0^\infty e^{-s} \frac{\pa H_1}{\pa x}(0, \nu s)[e^{-\lambda_2 s}-1]ds,
\end{align*}
which we combine with Equation \eqref{eq:g_nu2} to yield two independent eigenvalue equations,
\begin{align*}
 \lambda_1 &= -\frac{\nu}{\int_0^\infty e^{-s} H_1(\nu s, 0) ds} \int_0^\infty e^{-s} \frac{\pa H_1}{\pa x}(\nu s, 0)[e^{-\lambda_1 s}-1]ds,\\
 \lambda_2 &= -\frac{\nu}{\int_0^\infty e^{-s} H_1(\nu s, 0) ds} \int_0^\infty e^{-s} \frac{\pa H_1}{\pa x}(0, \nu s)[e^{-\lambda_2 s}-1]ds.
\end{align*}
Using these equations, we may determine stability of a traveling bump solution as a function of its velocity. We rephrase this problem and consider the independent scalar valued functions
\begin{align}\label{eq:Lambdai}
 \Lambda_1(\nu,\lambda) &= \lambda + \frac{\nu}{\int_0^\infty e^{-s} H_1(\nu s, 0) ds} \int_0^\infty e^{-s} \frac{\pa H_1}{\pa x}(\nu s, 0)\left(e^{-\lambda s}-1\right)ds,\\
 \Lambda_2(\nu,\lambda) &= \lambda + \frac{\nu}{\int_0^\infty e^{-s} H_1(\nu s, 0) ds} \int_0^\infty e^{-s} \frac{\pa H_1}{\pa x}(0, \nu s)\left(e^{-\lambda s}-1\right)ds.
\end{align}
For a given $\Lambda_i$, the zero level curves in $(\nu,\lambda)$ space determine stability properties of traveling bump solutions.

\begin{figure}[h!]
\centering
\includegraphics[width=\textwidth]{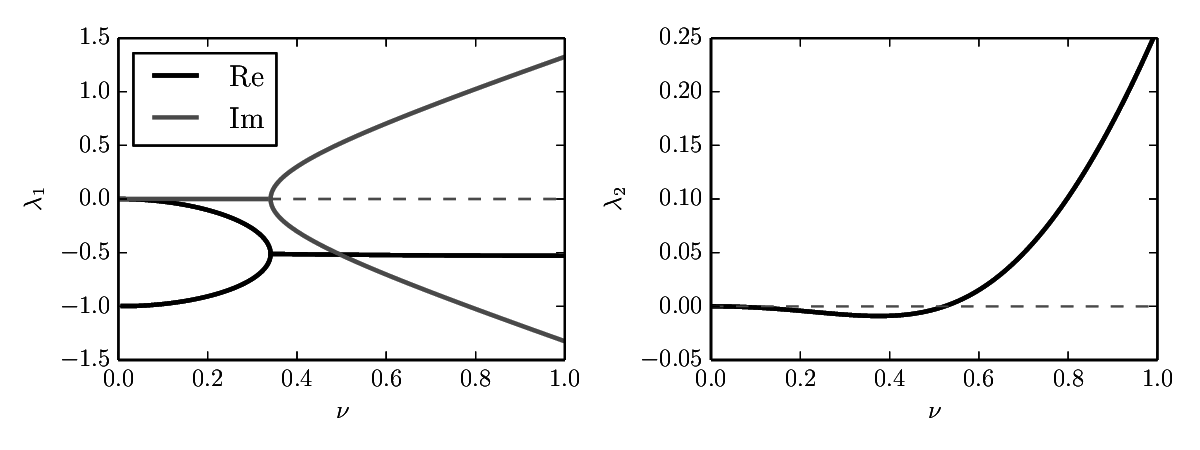}
 \caption{\ypre{Stability of solutions in the horizontal axial direction (calculated using the phase model with the accurate Fourier series). Both plots show the level curves where $\text{Re}(\Lambda_i)=0$ (black) and $\text{Im}(\Lambda_i)=0$ (gray). For small velocities $\nu$, both components are stable. For larger velocities, the horizontal velocity remains stable, but the vertical velocity loses stability. The dashed gray line shows where $\lambda_i = 0$.}}\label{fig:axial_stability}
\end{figure}

\ypre{We begin the analysis of these equations using the accurate Fourier series of $H_i$ and compute the integrals explicitly. The zero level set of the resulting function is shown in Figure \ref{fig:axial_stability}. On the left panel, find that for any velocity, the $x$-direction is always stable. On the right, we find that for sufficiently small velocities, $\theta_2(\tau) = 0$ is a stable solution. Thus, constant velocity traveling solutions in this parameter regime will converge to the $x$-axis. Finally, for greater traveling bump velocities, the vertical direction loses stability, giving rise to non-axial solutions.
}

With the truncated $H_i$, which we recall to be $H_i^F(\theta_1,\theta_2) = \sin(\theta_1)(1+b\cos(\theta_2))$, we may compute the equations $\Lambda_i=0$ explicitly as polynomials,
\begin{align*}
 0 &= \lambda _1^2+\lambda _1+2 \nu ^2\\
 0 &= \lambda _2^3+c_2\lambda _2^2+c_1\lambda_2 + c_0,
\end{align*}
where
\begin{equation}\label{eq:ci}
 \begin{split}
 c_0 &= \frac{(2b-1)\nu^2 - \nu^4}{b+1},\\
 c_1 &= \frac{\left(1+b + (2b-1)\nu^2\right)}{b+1},\\
 c_2 &= \frac{\left(2 (b+1)-\nu ^2\right)}{b+1}.
 \end{split}
\end{equation}

\ympree{The coefficients of Equation \eqref{eq:ci} determine the stability of the horizontal traveling solution. Note that for $\nu$ sufficiently small, all coefficients are positive and the product $c_1c_2$ dominates the coefficient $c_0$. Thus, for small velocities, the coefficients have the properties $c_1c_2>c_0$ and $c_2,c_0>0$, which implies stability by the Routh Hurwitz criterion. When $\nu^* = \pm\sqrt{2b-1}$, the coefficient $c_0$ is no longer positive and the stability condition fails.

We have found that horizontal traveling bump solutions lose stability at some critical velocity $\nu^*$, giving rise to non-axial traveling bump solutions. By symmetry, this argument holds for vertical traveling bump solutions: after the same critical $\nu^*$, constant velocity traveling bumps in the vertical direction lose stability and become non-axial solutions.

Now that we understand the existence of and stability of axial constant velocity traveling solutions, we turn our attention to the stability of non-axial traveling bump solutions.}

\ypre{
To determine the stability of non-axial directions, we consider the solution, $\theta_1(\tau) = \nu \tau + \phi_1 e^{\lambda\tau}$ and $\theta_2(\tau) = \nu \tau + \phi_2 e^{\lambda\tau}$. This ansatz results in the equations,
\begin{align*}
 \lambda \phi_1 &= -g \phi_1\int_0^\infty Q_1(s)\left(e^{-\lambda s}-1\right)ds-g \phi_2\int_0^\infty Q_2(s)\left(e^{-\lambda s}-1\right)ds,\\
 \lambda \phi_2 &= -g \phi_1\int_0^\infty Q_3(s)\left(e^{-\lambda s}-1\right)ds-g \phi_2\int_0^\infty Q_4(s)\left(e^{-\lambda s}-1\right)ds,
\end{align*}
where
\begin{align*}
 Q_1(s) = \int_0^\infty e^{-s} \frac{\pa H_1}{\pa x}(-\nu_1 s,-\nu_2 s), &\quad Q_2(s) = \int_0^\infty e^{-s} \frac{\pa H_1}{\pa y}(-\nu_1 s,-\nu_2 s),\\
 Q_3(s) = \int_0^\infty e^{-s} \frac{\pa H_1}{\pa y}(-\nu_2 s,-\nu_1 s), &\quad Q_2(s) = \int_0^\infty e^{-s} \frac{\pa H_1}{\pa x}(-\nu_2 s,-\nu_1 s).
\end{align*}
By rewriting the integrals in the more compact form,
\begin{align*}
 \lambda \phi_1 &= -g \phi_1 \hat Q_1(\lambda)-g \phi_2\hat Q_2(\lambda),\\
 \lambda \phi_2 &= -g \phi_1 \hat Q_3(\lambda)-g \phi_2\hat Q_4(\lambda),
\end{align*}
where $\hat Q_i = \int_0^\infty Q_i(s) (e^{-\lambda s} -1)ds$, the problem reduces to finding an eigenvector $(\phi_1,\phi_2)^T$ with corresponding eigenvalue $-\lambda$:
\begin{equation}\label{eq:eigenvalue_evans}
g\left(\begin{matrix}
 \hat Q_1(\lambda) & \hat Q_2(\lambda)\\
 \hat Q_3(\lambda) & \hat Q_4(\lambda)
\end{matrix}\right)
\left( \begin{matrix}
        \phi_1\\ \phi_2
       \end{matrix}
\right)
 =-\lambda\left( \begin{matrix}
        \phi_1\\ \phi_2
       \end{matrix}
\right).
\end{equation}
This condition holds if and only if the determinant
\begin{equation}\label{eq:evans}
\mathcal{E}(\lambda) = \left|g\left(\begin{matrix}
 \hat Q_1(\lambda) & \hat Q_2(\lambda)\\
 \hat Q_3(\lambda) & \hat Q_4(\lambda)
\end{matrix}\right) + \lambda I_2 \right|
\end{equation}
is zero. \ympree{This determinant is the Evans function, and we use its roots to determine stability properties of the constant velocity solutions.

Using the accurate Fourier series of $H_i$, the integrals of the eigenvalue problem \eqref{eq:eigenvalue_evans} are explicitly computable. Given a value $g$, it is straightforward to compute the contours of $\mathcal{E} = 0$ using a standard contour plot routine. In panels A and B of Figure \ref{fig:evans_full}, we show the Evans function when $g=1.5$, and $g=3$, respectively.

\begin{figure}[h!]
    \centering
\includegraphics[width=.9\textwidth]{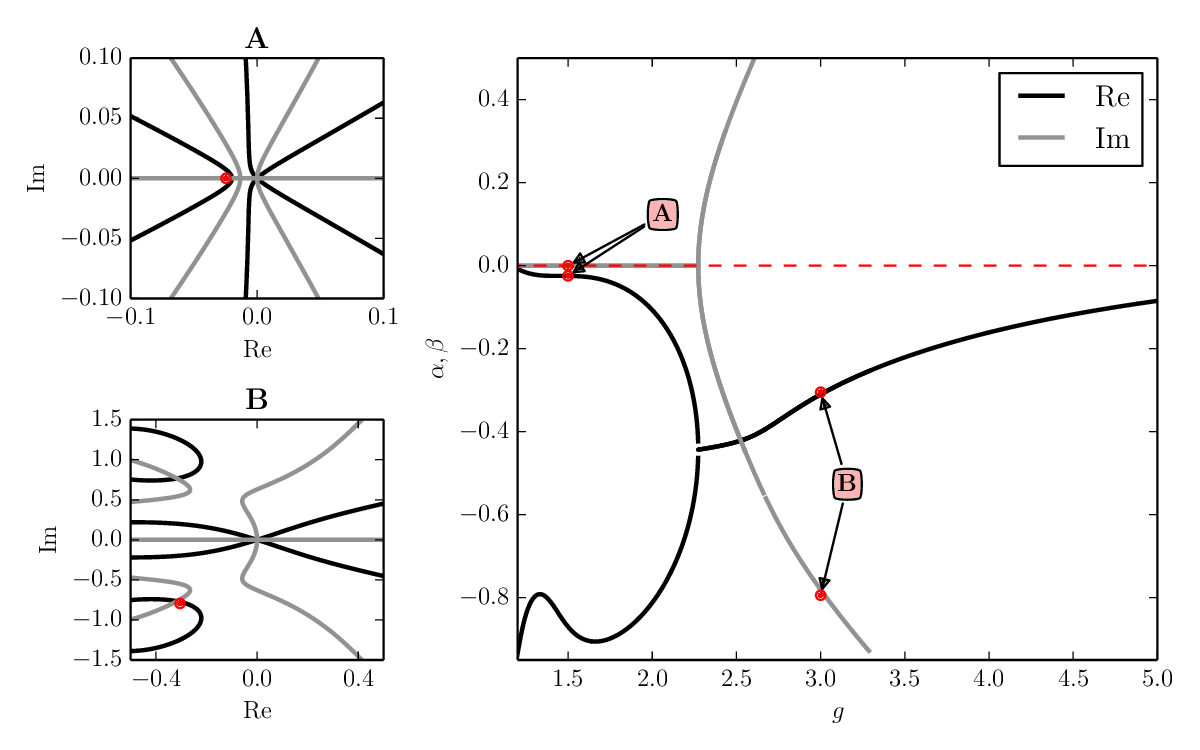}
    \caption{\ypre{Evans function for the accurate Fourier series $H$. Left panels A and B: roots of the real (black) and imaginary (gray) parts of the evans function for $g=1.5$ and $g=3$, respectively. Intersections of the gray and black lines denote zeros of the Evans function. We use $\alpha$ and $\beta$ to denote the real and imaginary part of the Evans function, respectively. Right panel: The real and imaginary parts of the nontrivial root(s) of the Evans function for various choice of $g$. The horizontal red dashed line denotes the real axis.}}\label{fig:evans_full}
\end{figure}

The domain values where the real part of the Evans function is zero is shown as a black contour, while the domain values where the imaginary part is zero is shown in gray. Intersections of these contours show roots of the Evans function. Generally, there exists a root of the Evans function at the origin due to translation invariance of the underlying bump solution. Thus we ignore this root and consider only those nontrivial roots with real part sufficiently greater than $-1$. These nontrivial roots are marked with red dots.

We follow these roots using \texttt{XPPAUTO} and generate the bifurcation diagram shown in the right panel of Figure \ref{fig:evans_full}. The real part of the root remains negative for the range of $g$ that we consider, thus the constant traveling bump solution remains stable for a large range of adaptation strengths.}

\ympree{We repeat the analysis of the Evans function using the truncated Fourier interaction function, $H^F$. Once again, the integrals of the eigenvalue problem \eqref{eq:eigenvalue_evans} are explicitly computable and we follow the roots of the Evans function using \texttt{XPPAUTO} in two parameters, $b$ and $g$, the Fourier coefficient, and adaptation strength, respectively. The right panel of Figure \ref{fig:evans_2par} shows the result of this continuation: within the unstable region (marked in light blue), constant velocity solutions are unstable, as demonstrated by the lower inset showing $\theta_1$ as a function of time. Because the instability arises through a Hopf bifurcation, the traveling bumps begin to ``wobble''. In the stable region, bump solutions travel with constant velocity, as demonstrated by the upper inset showing $\theta_1$ as a function of time.

\begin{figure}[h!]
    \centering
\includegraphics[width=.9\textwidth]{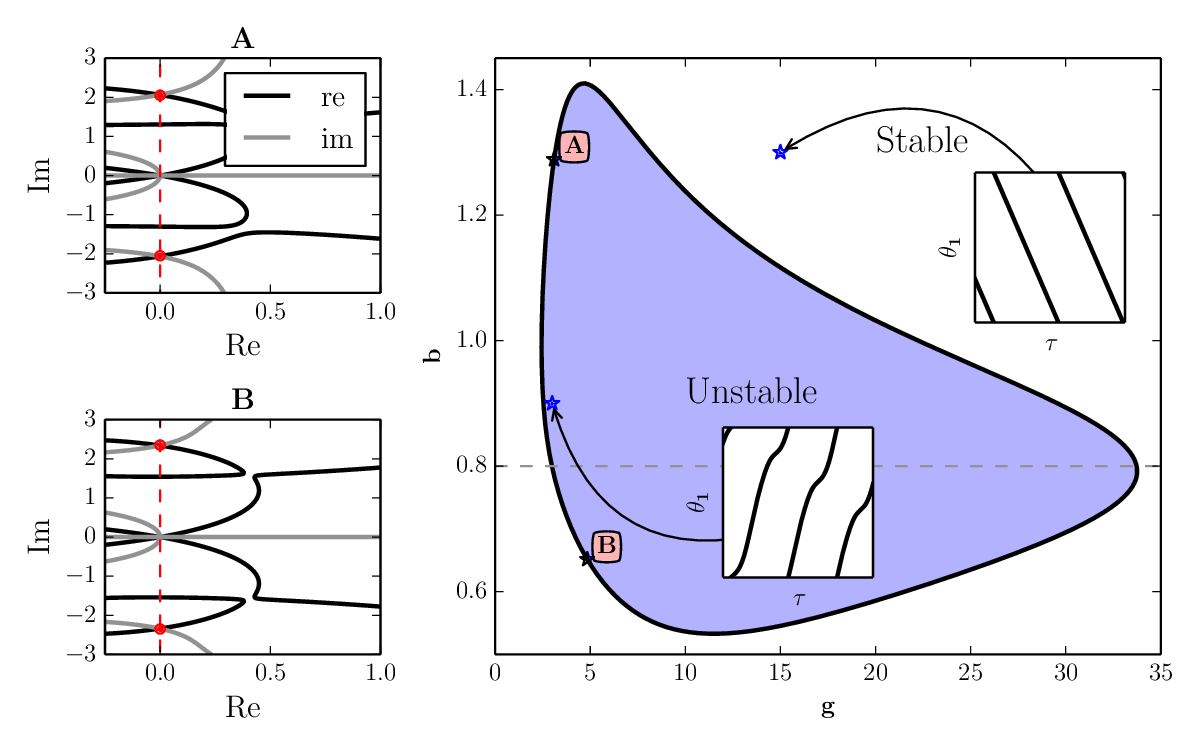}
    \caption{\ypre{Evans function for the truncated interaction function $H^F$. Left panels A and B: roots of the real (black, $g=3.1$, $b=2.28$) and imaginary (gray, $g=4.84$, $b=0.65$) parts of the evans function \ympre{demonstrating a loss of stability through a Hopf bifurcation}. Intersections of the gray and black lines denote zeros of the Evans function. Right panel: \ympree{the black line denotes where the real part of the Evans function is zero} in $b$ and $g$ parameter space (i.e., where the bump solution loses stability). The points labeled A and B correspond to panels A and B, respectively. The horizontal dashed gray line shows our usual choice of the parameter value $b=0.8$.\ympre{Two insets with example solutions of $\theta_1(\tau)$ over slow time $\tau$ are shown, corresponding to the blue star in parameter space. In the stable region, the traveling bump solution moves with constant velocity (inset parameter values $g=15$,$b=1.3$ integrated over $t=20000$ time units with the last 7.5\% of the data shown). In the unstable region, the traveling bump solution loses stability through a Hopf bifurcation and begins to travel with nonconstant velocity (inset parameter values $g=3$,$b=0.9$ integrated over $t=20000$ time units with the last 2.5\% of the data shown). Parameter: $\ve = 0.01$.}}}\label{fig:evans_2par}
\end{figure}

The left panels (A and B) of Figure \ref{fig:evans_2par} demonstrates the existence of a Hopf bifurcation on the boundary between stable and unstable regions. These panels correspond to points labeled A and B on the right panel. In each case, we find a complex conjugate pair of eigenvalues that cross the imaginary axis.}

}
In this section, we analyzed the reduced neural field model with nonzero adaptation strength ($g>0$) and no input current ($q=0$). We now explore the dynamics arising from activating the time-invariant input current.

\subsection{Hopf Bifurcation on the Torus}
\ypre{We have seen in Figures \ref{fig:2d2par}, \ref{fig:full_q=.1} that for nonzero $g$ and $q$, the system may produce a traveling bump solution that oscillates about the origin-centered input}. For a fixed parameter value $q$, the origin is stable for $g=0$, and with increasing $g$ eventually becomes unstable through a Hopf bifurcation.  We study this phenomenon with Equation \eqref{eq:phs_rscl2}, using the same technique as used on the ring: linearization about the origin.

\begin{figure}[h!]
    \centering
\includegraphics[width=\textwidth]{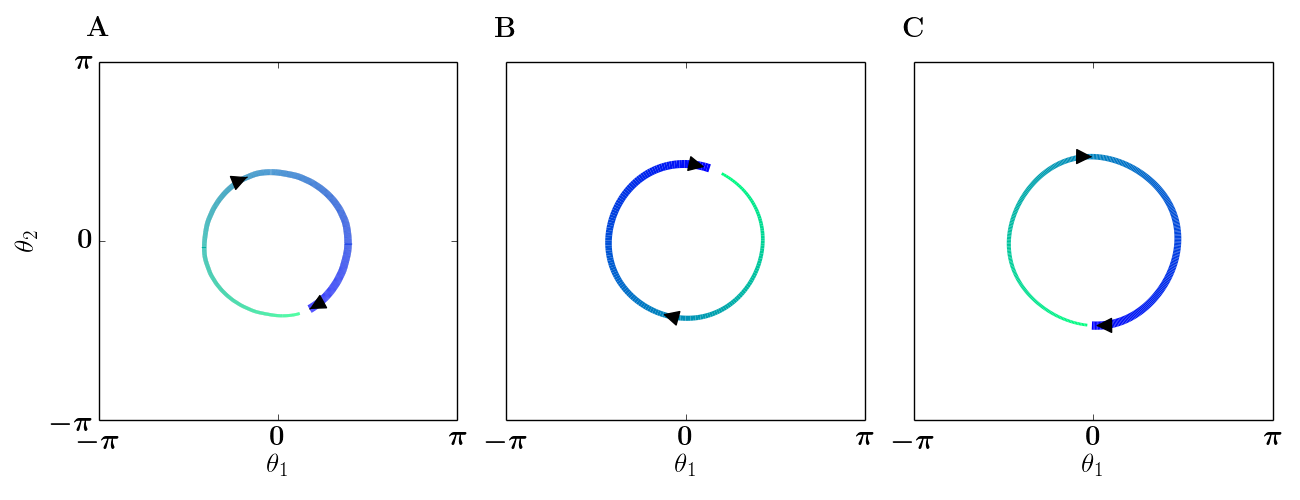}
    \caption{Limit cycle dynamics of the traveling bump on the torus. \ypre{The curve that goes from light to dark and thin to thick represents the movement of the centroid over time.} A: Full neural field model on the torus, $q=2$, $g=5$, \ypre{period of $t=805$ time units}. B: Phase model on the torus with the accurate Fourier series approximation of $H_i$, $q=1$, $g=3$, \ypre{period of $t=187$ time units}. C: Phase model on the torus with the truncated Fourier series $H_i^F$, $q=0.2$, $g=1$, \ypre{period of $t=370$ time units}. Parameter $\ve = 0.01$. }\label{fig:2d_wobble}
\end{figure}

Let $(\theta_1(\tau),\theta_2(\tau)) = (e^{\lambda \tau}, e^{\lambda \tau})$. Plugging into Equation \eqref{eq:phs_rscl2} results in a system of two decoupled equations,
\begin{equation*}
 \lambda = q \hat J_i^0 - g \hat H_i^0 \int_0^\infty e^{-s}(e^{-\lambda s}-1) ds,
\end{equation*}
where
\begin{align*}
 \hat J_i^0 &= \frac{\pa J_i}{\pa x}(0,0) + \frac{\pa J_i}{\pa y}(0,0),\\
 \hat H_i^0 &= \frac{\pa H_i}{\pa x}(0,0) + \frac{\pa H_i}{\pa y}(0,0).
\end{align*}
Evaluating the integral and solving for $\lambda$ yields
\begin{equation*}
 2\lambda = -(1 - q\hat J_i^0 - g\hat H_i^0) \pm \sqrt{(1 - q\hat J_i^0 - g\hat H_i^0)^2 + 4 q \hat J_i^0}.
\end{equation*}
Thus, as in the case of the ring, for a fixed $q$ and given $g$ sufficiently large, there exists a Hopf bifurcation at the critical value
\begin{equation*}
 g = \frac{1-q\hat J_i^0}{\hat H_i^0}.
\end{equation*}
For the truncated interaction function $H_i^F$, the critical value is
\begin{equation*}
 g = \frac{1+q(1+b)}{1+b}.
\end{equation*}

\subsection{Non-Constant Velocity Bump Solution on the Torus}
\begin{figure}[htbp]
    \centering
        \includegraphics[width=\textwidth]{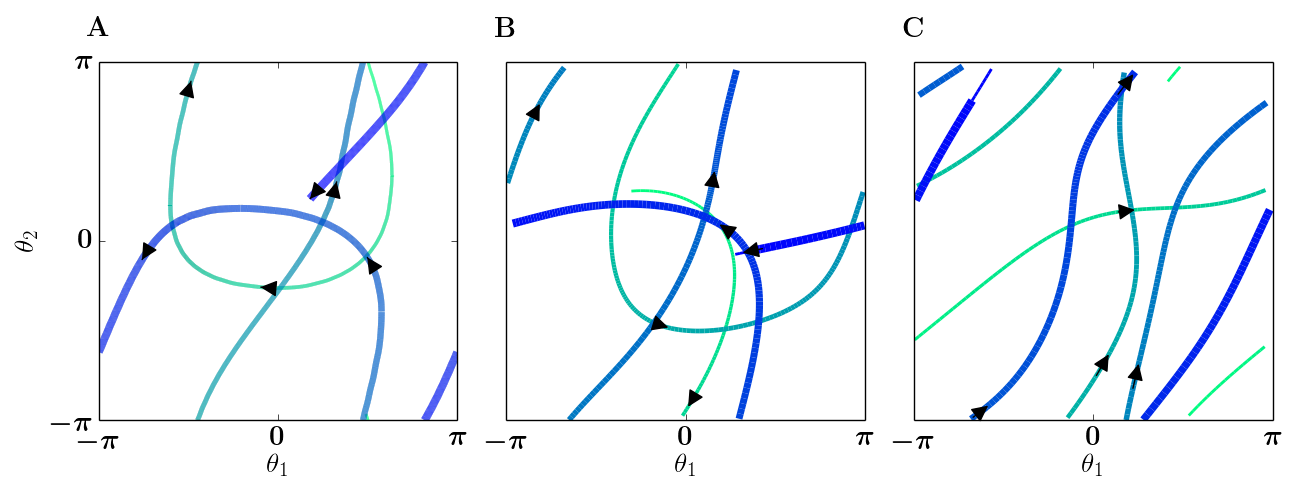}
    \caption{Non-constant velocity dynamics of the traveling bump on the torus. \ypre{The curve that goes from light to dark and thin to thick represents the movement of the centroid over time.} A: Full neural field model on the torus, $q=1$, $g=5$, \ypre{simulated for $t=5,000$ time units with the last 60\% of the data shown}. B: Phase model on the torus with the accurate Fourier series of $H_i$, $q=1$, $g=5$, \ypre{simulated for $t=6,700$ time units with the last 8\% of the data shown}. C: Phase model on the torus with the truncated Fourier series $H_i^F$, $q=.5$, $g=4.5$, \ypre{simulated for $t=6,700$ time units with the last 7\% of the data shown. Parameter: $\ve = 0.01$.}}\label{fig:2d_nonconst}
\end{figure}

As we have seen in earlier sections, stable oscillating solutions exist for particular choices of input current strength and adaptation on both the ring and torus. The similarities of solutions on the ring and torus continue as adaptation strength increases. On the ring, the oscillating solution gives way to a bump solution that travels around the ring with non-constant velocity. Similarly, with sufficiently large adaptation $g$, the bump solution on the torus also breaks free from the oscillating solution and traverses the domain with non-constant velocity. Figure \ref{fig:2d_nonconst} shows examples of these solutions in the full model (panel A), the phase model with the accurate Fourier series (panel B), and the phase model with the truncated Fourier series (panel C).

There are plenty of other examples of these types of solutions (Figures \ref{fig:q=.1},\ref{fig:2dfull2par},\ref{fig:full_q=.1}) that are in fact chaotic. To demonstrate the existence of chaos numerically, we use the truncated phase model and a Poincar\'{e} section through $cy=0$, as we find that generically the variable $cy$ consistently crosses zero throughout simulations.

\subsubsection{Chaos on the Torus}
\begin{figure}[htbp]
    \centering
        \includegraphics[width=.75\textwidth]{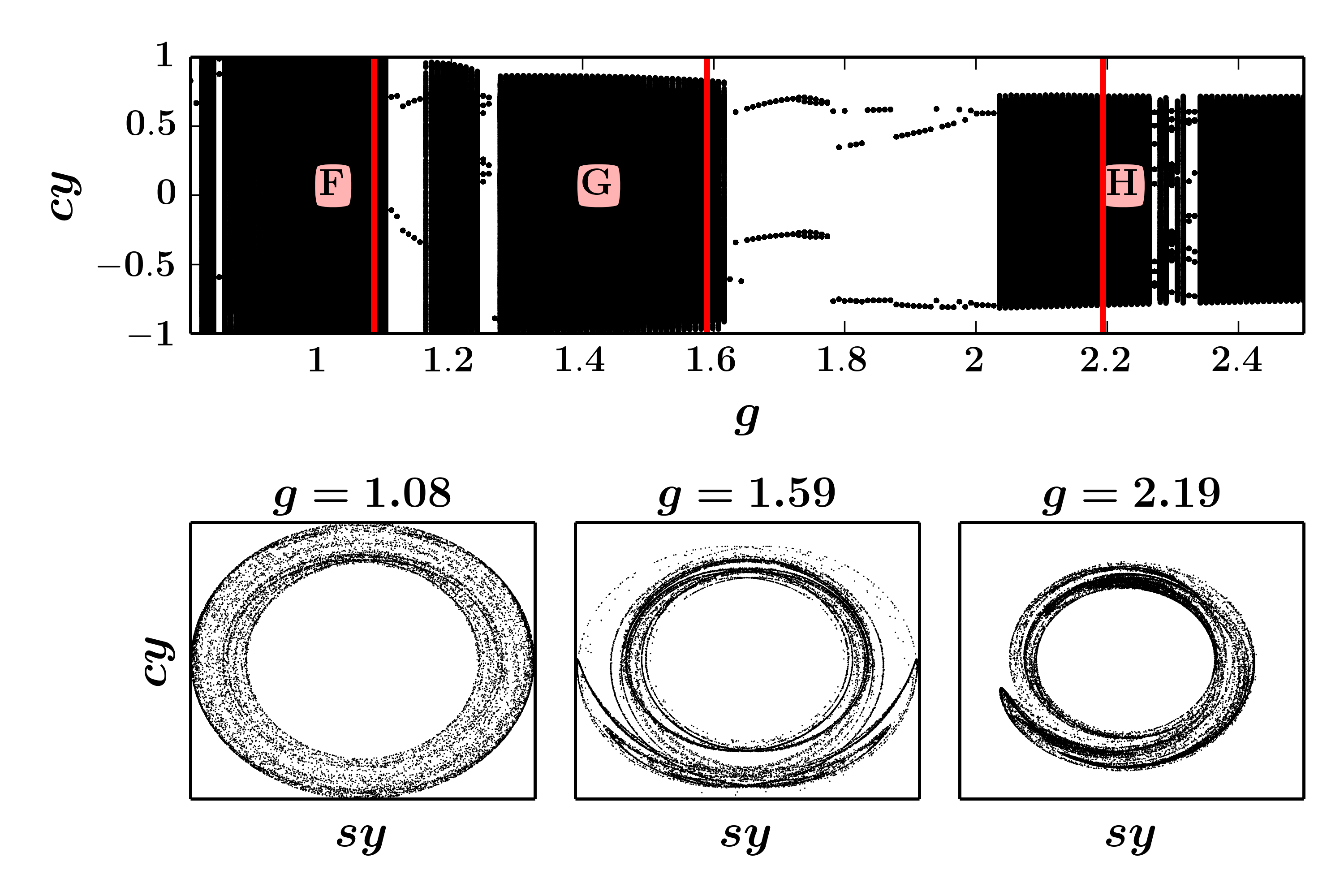}
    \caption{Chaotic attractors. Top panel: crude bifurcation diagram of $cy$ as a function of parameter $g$. Black regions correspond to aperiodic and possibly chaotic behavior, while regions with dots correspond to periodic solutions. Parameter: $q=0.1$.}\label{fig:2d_chaos}
\end{figure}

For a given $g$, we simulate the truncated phase model (Equation \eqref{eq:phase_trunc}) for $t=150000$ time units and ignore the first $7000$ time units to remove transients. By plotting the appropriate state variables, we are able to determine whether a system is aperiodic (and possibly chaotic) or periodic. The top panel of Figure \ref{fig:2d_chaos} shows one example of one such plot, where for each $g$ value we plot all $cy$ values for the duration of the simulation. The black regions of Figure \ref{fig:2d_chaos} correspond to the gray regions of Figure \ref{fig:q=.1}: the approximate range $0.85 < g < 1.1$ corresponds to region \textbf{F}, the approximate range $1.18 < g < 1.61$ corresponds to region \textbf{G}, and the approximate range $g > 2.05$ corresponds to region \textbf{H}, respectively.

We show sample solutions of the chaotic attractors in regions \textbf{F},\textbf{G},\textbf{H} in the bottom three panels of Figure \ref{fig:2d_chaos}. The vertical red line in the top panel denotes the parameter value corresponding to each attractor.

%

\section{Discussion}
\be{Our motivation for this work was to understand the behavior of the model presented in \cite{itskov_2011_cell} where the authors showed that heterogeneities in a recurrent network with adaptation produced a seemingly randomly moving bump of activity.  Similar moving bump dynamics was also found in a homogeneous bump model with adaptation in \cite{fung2010}; here the authors report only axially moving bumps with no external inputs. }
\ympre{The neural field model considered in this paper (Equations \eqref{eq:u1},\eqref{eq:z1}) is also capable of producing a rich variety of solutions on the ring and torus. On the ring, the centroid of the bump solution exhibits sloshing and large-sloshing behaviors (for moderate strengths of input current and adaptation) that, for stronger adaptation, lead to non-constant velocity traveling bump solutions. With no input current and sufficient adaptation, the system generates a constant velocity traveling bump solution. We also observe chaotic solutions for a narrow range of adaptation strengths.

On the torus, the qualitative solutions are similar to those on the ring: Stationary bump solutions give rise to sloshing solutions (for moderate strengths of input current and adaptation), as well as non-constant velocity traveling bump solutions (with sufficient adaptation) and constant velocity traveling solutions (with no input current and sufficient adaptation). In this system, we do not see pulses that change in diameter with a fixed centroid (breathers) on the ring or torus.

Neural fields with nonsmooth firing rate functions (i.e., the Heaviside or rectifying nonlinearity) reproduce many of these qualitative behaviors. The existence and bifurcation of sloshing solutions on the ring are analyzed in \cite{folias2011nonlinear,ermentrout2014spatiotemporal}, and constant velocity solutions are shown to exist on the ring (\cite{york_recurrent_2009,ermentrout2014spatiotemporal}), the real line (\cite{kilpatrick_effects_2010}), and the plane (\cite{fung_spontaneous_2015}). Nonconstant velocity bump solutions are shown to exist in \cite{ermentrout2014spatiotemporal}. However, there are no studies showing the existence of aperiodic attractors on the torus (assuming a deterministic system with an even kernel), or the existence of chaos on the ring.

In this study, we contributed to the analysis of the known behaviors by using a smooth firing rate function and a caveat of weak and slow adaptation. This assumption on the adaptation variable allowed us to reduce the neural field model, which is a distributed partial integro-differential equation, to a system of scalar delay integro-differential equations describing the centroid of the bump solution. Moreover, our only restriction on the kernel is of Mexican-hat type. Put together, these assumptions and our results are more general than what currently exists in literature. 

In one spatial dimension, for example, we derived the normal form for the Hopf bifurcation in the one-dimensional neural field model and determined the conditions for super- and sub-criticality. Although normal form calculations exist for neural field models on the ring or the real line, our calculation allows for a general choice of kernel and a smooth firing rate function (as opposed to a particular choice of kernel or a non-smooth Heaviside firing rate function \cite{ermentrout2014spatiotemporal,folias2011nonlinear}).}

As mentioned previously, existing studies require particular choices of kernels or Heaviside firing rate functions where a smooth firing rate function would be desirable (\cite{bressloff_folias_2004_siam} note that a smooth firing rate function allows for a straightforward normal form analysis). Although these assumptions are restrictive, these studies have advantages that the current study does not address. In particular, our analysis requires that adaptation is weak and slow and that the input current is weak. As a result, we can only study phenomena that evolve on a slow timescale. These weak and slow assumptions are well-suited for studying long-lasting sequences of spatially coherent activity in the absence of changing external stimuli \cite{pastalkova_2008_science,itskov_2011_cell}, but may not be as well suited to study phenomena on a faster timescale, like the effects of weak modulatory interactions mediated by the reciprocal, long-range \textit{patchy connections} in primary visual cortex \cite{folias2011nonlinear}.

\be{ Generally speaking, one might ask why we need the adaptation to be both slow and weak. For example, in \cite{pinto_ermentrout_2001_siam}, the adaptation was slow but not weak.  One could imagine doing a perturbation analysis for a weak stimulus such as in \cite{ermentrout2010stimulus} where a weak slowly moving stimulus is applied to a system that has a stable traveling bump. However, strong adaptation, will alway induce movement in a bump so that we can never pin the bump with {\em weak} inputs. Furthermore, by keeping the adaptation $O(1)$, one needs to compute the adjoint solution to a two-variable traveling bump, a difficult task in one spatial dimension, and impossible (as far as we can tell) in two spatial dimensions.  Thus, by working with weak inputs and weak/slow adaptation, we have hit a sweet spot from which many of the interesting dynamics emerges.} 

\bge{One type of behavior that has been observed in this class of models that does not occur in our analysis is the so-called breathing solutions \cite{folias_bressloff_2004_siam}. Breathers are periodic solutions to the neural field equations that occur when the bump solution loses stability via a {\em symmetric} mode. In contrast, sloshers appear when there is a Hopf bifurcation to an {\em anti-symmetric} mode.  In the breather case, the centroid of the bump does not change, so our reduced equations cannot detect such a bifurcation. In contrast, sloshers lead to modulation of the centroid and thus our analysis can capture that. To further explore this, we were able to induce bifurcation to a breathing solution in equations (\ref{eq:u1},\ref{eq:z1}) but only when $\varepsilon$ is sufficiently large. We find that it is possible to continue this bifurcation in $\varepsilon$ and make $\varepsilon$ quite small, but only if we increase both the strength of adaptation $g$ and the heterogeneity, $q$ such that $\varepsilon q, \varepsilon g$ remain $O(1).$ That is, breathers can only occur when the adaptation and input magnitudes are large compared to the rate of adaptation. Our analysis, therefore, cannot include the appearance of breathers.}

\ymp{The effects of noise on the phase equations is one possible direction for future study. Several cited papers analyze the movement of bump solutions in the presence of noise. Near the drift bifurcation for traveling bump solutions with sufficiently strong linear adaptation, it is possible to derive a stochastic amplitude equation when the adaptation strength and stochastic forcing are similar in magnitude \cite{kilpatrick_pulse_2014}. Sufficiently far from the bifurcation, the stochastic forcing leads to diffusive wandering of the bump solution. Other studies analyze the diffusive behavior of solutions to neural field models and how pinning eliminates diffusive behavior \cite{poll_stochastic_2015}. In \cite{laing_noise-induced_2001}, the authors compute the effects of adding noise to the normal form of the pitchfork bifurcation. As mentioned previously, these studies assume either a particular firing rate function or kernel. The general case remains unexplored.}

\appendix

\section{Normal Form for the Hopf Bifurcation on the Ring}\label{a:normal}
Recall that we analyze the bifurcation to sloshing pulses for the general integral equation,
\begin{equation}
\frac{d\theta}{d\tau} = -q J(\theta) - g \int_0^\infty e^{-s} H(\theta(\tau-s)-\theta(\tau))\ ds
\end{equation}
as $g$ increases.  For simplicity, we assume the expansions
\begin{eqnarray*}
J(\theta) &=& \theta + j_3 \theta^3 + \ldots \\
H(\theta) &=& \theta + h_3 \theta^3 + \ldots,
\end{eqnarray*}
and $q>0$. Based on the eigenvalue equation,
\[
\lambda^2 +(1+q-g)\lambda+ q=0
\]
we expect a Hopf bifurcation to occur.

To analyze the Hopf bifurcation, we use a multiple time scale expansion.  We assume that $\theta(\tau)$ is a function of a ``fast'' time $\zeta=\tau$ and a ``slow'' time $\xi=\delta^2\tau$ where $\delta$ measures the amplitude of the bifurcating solution. As the nonlinearities are all odd, we can assume that 
\[
g= g_0 + \delta^2 g_2, \qquad \theta = \delta \theta_1(\zeta,\xi) + \delta^3 \theta_3(\zeta,\xi)
\]
to order $\delta^3$. We develop a perturbation expansion to obtain the normal form. Before continuing, we need to briefly describe how the integral equation gets expanded in multiple scales. If $f(\zeta,\xi)$ is a function of the fast and slow time-like variable, then, clearly
\[
\frac{df}{d\tau} = \frac{\partial f}{\partial \zeta} + \delta^2 \frac{\partial f}{\partial \xi} 
\]
and 
\[
\int_0^\infty e^{-s} f(\tau-s) \ ds =  \int_0^\infty e^{-s} f(\zeta-s,\xi - \delta^2 s) \ ds.
\]
We expand this expression to order $\delta^2$ to get:
\begin{equation}
\label{eq:slowint}
 \int_0^\infty e^{-s} f(\tau-s) \ ds \approx \int_0^\infty e^{-s} f(\zeta-s,\xi) \ ds  - \delta^2 \int_0^\infty s e^{-s} \frac{\partial f(\zeta-s,\xi)}{\partial \xi} \ ds.
\end{equation}
Let 
\[
(L u)(\zeta):= \frac{\partial u}{\partial \zeta} + q u + g_0 \int_0^\infty e^{-s} [u(\zeta-s)-u(\zeta)] \ ds.
\]
By our choice of $g_0$, $L$ has a nullspace $e^{\pm i\omega\zeta}$ and since it is a scalar, so does the adjoint operator under the usual inner product
\[
\langle u,v\rangle:= \int_0^{2\pi/\omega} \bar{u}(s)v(s)\ ds.
\]  
We plug in all the expansions and find to first order that
\[
\theta_1 = z(\xi) e^{i\omega \zeta} + c.c
\]
where $z(\xi)$ is a complex function of $\xi$ and c.c means complex conjugates.  Our goal is to derive equations for $z$.  To cubic order, we obtain:
\begin{align*}
(L \theta_3)(\zeta)&=  z_\xi e^{i\omega\zeta}\left(-1 + \frac{g_0}{1 + 2 i \omega -\omega^2}\right) + c.c\\
 &+  g_2 z e^{i\omega\zeta}\frac{i\omega}{1+i\omega} + c.c \\
 &+  -q j_3 \left[ z e^{i\omega \zeta} + \bar{z}e^{-i\omega\zeta}\right]^3  \\
 &+ -g h_3 \int_0^\infty \left[z(\xi)e^{i\omega\zeta}(e^{-i\omega s}-1) + \bar{z}(\xi)e^{-i\omega\zeta}(e^{i\omega s}-1)\right]^3 \ ds.
\end{align*}  
The first line comes from applying equation (\ref{eq:slowint}). Taking the inner product of this equation with $\exp(i\omega\zeta)$ (essentially, the Fredholm alternative), yields the equation for $z(\xi)$:
\begin{equation}
\alpha \frac{d z}{d\xi} = z [\hat{\gamma}_0 + \hat{\gamma}_3 |z|^2]
\end{equation}
where
\begin{align*}
\alpha &= 1 - \frac{g_0}{1+2i\omega-\omega^2} = \frac{2}{1+q}( q + \sqrt{q}i) \\
\hat{\gamma}_0 &= g_2 \frac{i\omega}{1+i\omega} = \frac{g_2}{1+q}(q + \sqrt{q}i) \\
\hat{\gamma}_3 &= \frac{3q}{4q+1}\left[[q(12h_3-4j_3)-j_3]+i 18h_3 \sqrt{q}\right].
\end{align*}

\section{Computation of Functions \texorpdfstring{$H_i$}{TEXT} and \texorpdfstring{$J_i$}{TEXT}}
To numerically integrate the phase models on the ring or torus, we require an approximation to the functions $H_i$, and $J_i$. These functions depend on and use lookup tables for the steady state bump $u_0$ (\texttt{u0ss}), the derivative of the firing rate evaluated at the steady state bump $f'(u_0)$ (\texttt{df\_u0b}), and the partial derivatives of the steady state bump, $\partial u_0/\partial x$, $\partial u_0/\partial y$ (\texttt{ux},\texttt{uy}). On the toroidal domain, each lookup table has $N\times N$ entries, where for the coefficients below, we choose $N=64$.

To compute $H_i$ in Equation \eqref{eq:theta2},we use the following procedure

\begin{verbatim}
 H1 = zeros(N,N)
 H2 = zeros(N,N)
 for i=1:N
   for j=1:N
     temp1 = 0
     temp2 = 0
     for n=1:N
       for m=1:N
         xn = mod(n+i+N/2,N)
         xm = mod(m+j+N/2,N)
         temp1+=ux[n,m]*df_u0b[n,m]*u0ss[xn,xm]
         temp2+=uy[n,m]*df_u0b[n,m]*u0ss[xn,xm]
       end
     end
     H1[i,j] = temp1
     H2[i,j] = temp2
   end
 end
 H1 *= (2*pi)^2/N^2
 H2 *= (2*pi)^2/N^2
\end{verbatim}

To compute $J_i$ in Equation \eqref{eq:theta2}, we use the following procedure

\begin{verbatim}
 J1 = zeros(N,N)
 J2 = zeros(N,N)
 for i=1:N
   for j=1:N
     temp1 = 0
     temp2 = 0
     for n=1:N
       for m=1:N
         xn = mod(n+i+N/2,N)
         xm = mod(m+j+N/2,N)
         temp1+=ux[xi,xj]*df_u0b[xn,xm]*I[n,m]
         temp2+=uy[xi,xj]*df_u0b[xn,xm]*I[n,m]
       end
     end
     J1[i,j] = temp1
     J2[i,j] = temp2
   end
 end
 J1 *= (2*pi)^2/N^2
 J2 *= (2*pi)^2/N^2
\end{verbatim}

On the torus, taking the difference $J_i - (-H_i)$ results in a negligible error, revealing that $J_i = -H_i$. Thus, for all phase computations involving $J_i$, we use the same Fourier approximations for $H_i$ and $J_i$.

On the ring, the computations are virtually identical with the obvious exception of array shapes.

\subsection{Fourier Approximations}

After creating the lookup tables \texttt{H1,H2}, we perform a Fourier approximation to make numerical integration easier. The following function and corresponding coefficients and frequencies (Table \ref{tab:hi_fourier}) provide an excellent approximation to the lookup tables \texttt{H1,H2}. A basic error analysis shows that the supremum norm difference between the lookup tables \texttt{H1,H2} and their Fourier approximations,$\overline H_1, \overline H_2$, is $\| \texttt{H1}-\overline H_1 \|_\infty \approx 3.354\mathrm{e}{-7}$.
\begin{equation}\label{eq:fourier_2d}
 \overline H_1(x,y) = -\sum_{k=1}^{26}\frac{a_k}{N^2}\sin( x n_k + y m_k).
\end{equation}

\begin{table}
\caption {Fourier Coefficients of \texttt{H1} for $N=64$. The maximum pointwise difference between this approximation of $H_1$ and the original $H_1$ is \texttt{3.53733478176e-07}} \label{tab:hi_fourier}
\begin{center}
\begin{tabular}{l|l|l}
$k$ & $a_k$ & $(n_k,m_k)$\\
\hline
0  &  -0.299041640592  &  (1,0)\\
1  &  -0.0123427222227  &  (2,0)\\
2  &  -2.92404662557e-07  &  (3,0)\\
3  &  2.92404662711e-07  &  (-3,0)\\
4  &  0.0123427222227  &  (-2,0)\\
5  &  0.299041640592  &  (-1,0)\\
6  &  -0.110662059947  &  (1,1)\\
7  &  -0.00255677958311  &  (2,1)\\
8  &  -1.30119169782e-07  &  (3,1)\\
9  &  1.30119169839e-07  &  (-3,1)\\
10  &  0.00255677958311  &  (-2,1)\\
11  &  0.110662059947  &  (-1,1)\\
12  &  -0.00134078962566  &  (1,2)\\
13  &  -8.78193375763e-06  &  (2,2)\\
14  &  -1.40550932909e-07  &  (3,2)\\
15  &  1.40550932908e-07  &  (-3,2)\\
16  &  8.78193375764e-06  &  (-2,2)\\
17  &  0.00134078962566  &  (-1,2)\\
18  &  -0.00134078962566  &  (1,-2)\\
19  &  -8.78193375764e-06  &  (2,-2)\\
20  &  -1.40550932907e-07  &  (3,-2)\\
21  &  1.4055093291e-07  &  (-3,-2)\\
22  &  8.78193375763e-06  &  (-2,-2)\\
23  &  0.00134078962566  &  (-1,-2)\\
24  &  -0.110662059947  &  (1,-1)\\
25  &  -0.00255677958311  &  (2,-1)\\
26  &  -1.30119169783e-07  &  (3,-1)\\
27  &  1.30119169839e-07  &  (-3,-1)\\
28  &  0.00255677958311  &  (-2,-1)\\
29  &  0.110662059947  &  (-1,-1)
\end{tabular}
\end{center}
\end{table}

The coefficients in Table \ref{tab:hi_fourier} are computed using Python with Numpy by taking the Fourier transform of the lookup tables \texttt{H1,H2}.

\begin{table}
\caption {Fourier Coefficients of the steady-state coefficients. Plotting $u_{00} + 2u_{10}\cos(x) + 2u_{01}\cos(y) + 4u_{11}\cos(x)\cos(y)$ gives a reasonable approximation to the numerically computed steady-state bump solution.} \label{tab:k_fourier}
\begin{center}
\begin{tabular}{l|l|l}
$k$ & $u_k$ & $(n_k,m_k)$\\
\hline
0 & -2.17382490474 & (0, 0) \\
1 & -0.74563470929 & (0, 1) \\
5 & -0.74563470929 & (1, 0) \\
6 & 0.338867473649 & (1, 1) \\
7 & 0.340507108446 & (1, -1) \\
10 & -0.74563470929 & (-1, 0) \\
11 & 0.340507108446 & (-1, 1) \\
12 & 0.338867473649 & (-1, -1) \\
\end{tabular}
\end{center}
\end{table}

\begin{table}
\caption{Fourier Coefficients of the kernel. Plotting $k_{00} + 2k_{10}\cos(x) + 2k_{01}\cos(y) + 4k_{11}\cos(x)\cos(y)$ gives a reasonable approximation to the original periodix kernel.} \label{tab:u_fourier}
\begin{center}
\begin{tabular}{l|l|l}
$k$ & $k_k$ & $(n_k,m_k)$\\
\hline

0 & -0.473945684407 & (0, 0) \\
1 & 0.19095061386 & (0, 1) \\
4 & 0.19095061386 & (0, -1) \\
5 & 0.19095061386 & (1, 0) \\
6 & 0.108965377668 & (1, 1) \\
7 & 0.111033925698 & (1, -1) \\
10 & 0.19095061386 & (-1, 0) \\
11 & 0.111033925698 & (-1, 1) \\
12 & 0.108965377668 & (-1, -1) \\

\end{tabular}
\end{center}
\end{table}

\section{Numerical Integration}\label{a:numerical_int}

In this section, we detail the various numerical methods used to evaluate the many integro-delay-differential equations of this paper.

\subsection{Truncated Neural Field Model on the Torus}

The integration of Equation \eqref{eq:u_trunc_coeffs} requires the approximation of several double integrals. In the interest of reducing computation time, we use Riemann integrals and a relatively coarse discretization of the spatial domain. For example, for a given time $t$, the coefficient $p_{10}(t)$ is approximated as
\begin{equation*}
 p_{10}(t) \approx \sum_{n=1}^N \sum_{m=1}^N \cos(y_m) f(u(x_n,y_m,t)) \frac{(2\pi)^2}{N^2}.
\end{equation*}
Because a linear increase in $N$ leads to a quadratic increase in the total number of operations, we keep $N=100$, which is an acceptable compromise between speed and accuracy for this problem. All other double sums that appear in $p_{ij}$, $r_{ij}$, and $s_i$ are computed this way.

When computing the bifurcation diagram using this system, we use \texttt{XPPAUTO} and the numerical options shown in Table \ref{tab:xppauto_param_full}. The most important options are Ntst and Dsmin. If Ntst is less than 1000, \texttt{XPPAUTO} is unreliable in determining the stability of periodic solutions. If Dsmin is too large, \texttt{XPPAUTO} will skip bifurcation points.

\begin{table}
\caption{\texttt{XPPAUTO} parameters for the bifurcation diagram Figure \ref{fig:full_q=.1}. \texttt{XPPAUTO} version 8 has a third column of numerics options, which we left at default values.} \label{tab:xppauto_param_full}
\begin{center}
\begin{tabular}{l|l}
AUTO Option & Value \\
\hline
Ntst & 1000 \\
Nmax & 200 \\
NPr & 2  \\
Ds & 0.01 \\
Dsmin & 0.0001  \\
Ncol & 4  \\
EPSL & 0.0001  \\
Dsmax & 0.1  \\
Par Min & 0  \\
Par Max & 5  \\
Norm Min & 0  \\
Norm Max & 1000  \\
EPSU & 0.0001  \\
EPSS & 0.0001  \\
\end{tabular}
\end{center}
\end{table}

\subsection{Delay Integro-Differential Equations}
We implement the right hand side of the integro-differential in Equation \eqref{eq:phs_rscl2} as
\begin{equation*}
\begin{split}
 f\left(\begin{matrix}t_{k}\\ \vec x_M \\ \vec y_M\end{matrix}\right) &=\left(-g\left(\sum\limits_{n=0}^{M-1}e^{-n dt}H_1[x_{k-n}-x_{k},y_{k-n}-y_{k}]\right)dt+ qJ_1(x_{k},y_{k}) \right)\\
 h\left(\begin{matrix}t_{k}\\ \vec x_M \\ \vec y_M\end{matrix}\right) &=\left(-g\left(\sum\limits_{n=0}^{M-1}e^{-n dt}H_2[x_{k-n}-x_{k},y_{k-n}-y_{k}]\right)dt + qJ_2(x_{k},y_{k}) \right ),
\end{split}
\end{equation*}
where $dt$ is the time step and
\begin{equation*}
 \vec x_M = \left(\begin{matrix}
                  x_{k}\\ \vdots\\x_{k-(M-1)}
                  \end{matrix}
 \right),\quad
 \vec y_M = \left(\begin{matrix}
        y_{k}\\ \vdots\\y_{k-(M-1)}
       \end{matrix}
\right)
\end{equation*}
are the arrays containing solution values for $M$ previous time steps. The functions $H_i$ and $J_i$ are either the accurate Fourier approximation (Equations \eqref{eq:fourier_full1},\eqref{eq:fourier_full2}), or the truncated Fourier series (Equation \eqref{eq:h_trunc}). The time step $dt$ is the same as the discretization of the integral.

The algorithm is a straightforward Euler method. For a given time step \texttt{i},

\begin{verbatim}
 th1[i+1] = th1[i] + dt*f(t[i],th1[i],..,th1[i-(M-1)],
                              th2[i],..,th2[i-(M-1)])
 th2[i+1] = th2[i] + dt*h(t[i],th1[i],..,th1[i-(M-1)],
                              th2[i],..,th2[i-(M-1)])
\end{verbatim}
The initial condition for this algorithm requires an array of $M$ time steps. If the parameters are chosen such that a limit cycle exists, then we initialize in an arc:
\begin{verbatim}
 r0 = 1
 n0 = linspace(0,-.01,M)
 for k = 0:M-1
   th1[k] = r0*cos(n0[k])
   th2[k] = r0*sin(n0[k]).
\end{verbatim}
If the parameters are chosen such that a constant-velocity bump exists, then we initialize in a line:
\begin{verbatim}
 x_line = linspace(0,1,M)
 y_line = linspace(2,3,M)
 for k = 0:M-1
   th1[k] = x[k]
   th2[k] = y[k].
\end{verbatim}
When plotting solutions, we disregard at least the first $M$ entries of the solution vector.

\section*{Acknowledgments}
GBE would like to thank Rodia Curtu, Carina Curto, and Vladimir Itskov for early conversations about this problem.

\bibliographystyle{siamplain}
\bibliography{../ymp,../bio,../neuralfield}
\end{document}